\documentclass{aa}  

\usepackage[T1]{fontenc}
\usepackage{txfonts}
\usepackage{graphicx}
\usepackage[caption=false]{subfig}
\graphicspath{ {AP_plots/} }
\DeclareGraphicsExtensions{.pdf}
\usepackage{dcolumn} 
\usepackage{amsmath}
\usepackage{bm}

\usepackage{hyperref}
\hypersetup{
    colorlinks=true,
    linkcolor=blue,
    filecolor=magenta,      
    urlcolor=blue,
    citecolor=blue
}

\usepackage{xspace}

\newcommand{\voxel}{\texttt{voxel}\xspace}
\newcommand{\zobov}{\texttt{ZOBOV}\xspace}
\newcommand{\revolver}{\texttt{REVOLVER}\xspace}
\newcommand{\vide}{\texttt{VIDE}\xspace}
\newcommand{\pylians}{\texttt{Pylians3}\xspace}

\newcommand*{\superscr}[1]{\ensuremath{^{\rm #1}}}
\newcommand*{\subscr}[1]{\ensuremath{_{\rm #1}}}
\newcommand*{\pow}[1]{\ensuremath{\cdot 10^{#1}}}

\newcommand{\diff}{\ensuremath{\rm d}\xspace}
\newcommand{\hMpc}{\ensuremath{h^{-1}{\rm Mpc}}\xspace}

\newcommand{\aperp}{\ensuremath{q_\perp}}
\newcommand{\apar}{\ensuremath{q_\parallel}}
\newcommand{\astar}{\ensuremath{\alpha^*}}
\newcommand{\xir}{\ensuremath{\xi\superscr{rr}}}
\newcommand{\xis}{\ensuremath{\xi\superscr{rs}}}
\newcommand{\vecr}{\ensuremath{\mathbf{r}}}
\newcommand{\vecs}{\ensuremath{\mathbf{s}}}

\usepackage[nameinlink,noabbrev]{cleveref}
\crefname{equation}{Eq.}{Eqs.}
\crefname{section}{Sect.}{Sects.}
\crefname{figure}{Fig.}{Figs.}
\crefname{table}{Table}{Tables}
\crefname{appendix}{Appendix}{Appendices}
\Crefname{figure}{Figure}{Figures}
\Crefname{equation}{Equation}{Equations}
\Crefname{section}{Section}{Sections}
\Crefname{table}{Table}{Tables}

\begin{document} 

   \title{Alcock--Paczyński effect on void-finding:}
   \subtitle{Implications for void-galaxy cross-correlation modelling}
   \titlerunning{Alcock--Paczyński effect on void-finding}

\author{
	Sla{\dj}ana Radinović\inst{\ref{aff1},\ref{aff5}}\fnmsep\thanks{sladana.radinovic@astro.uio.no} \and 
	Hans A. Winther\inst{\ref{aff1}} \and
	Seshadri Nadathur\inst{\ref{aff2}} \and
	Will J. Percival\inst{\ref{aff3},\ref{aff4},\ref{aff6}} \and
	Enrique Paillas\inst{\ref{aff3},\ref{aff4}} \and
	Tristan Sohrab Fraser\inst{\ref{aff3},\ref{aff4}} \and
	Elena Massara\inst{\ref{aff3},\ref{aff4}} \and
	Alex Woodfinden\inst{\ref{aff3},\ref{aff4}}
}
\authorrunning{Radinović et al.}

	\institute{
	Institute of Theoretical Astrophysics, University of Oslo, PO Box 1029, Blindern 0315, Oslo, Norway\label{aff1} \and
	Institute of Space Sciences (ICE, CSIC), Campus UAB, Carrer de Can Magrans, s/n, 08193 Barcelona, Spain\label{aff5} \and
	Institute of Cosmology and Gravitation, University of Portsmouth, Burnaby Road, Portsmouth, PO1 3FX, UK\label{aff2} \and
	Waterloo centre for Astrophysics, University of Waterloo, 200 University Ave W, Waterloo, ON N2L 3G1, Canada\label{aff3} \and
	Department of Physics and Astronomy, University of Waterloo, 200 University Ave W, Waterloo, ON N2L 3G1, Canada\label{aff4} \and
    Perimeter Institute for Theoretical Physics, 31 Caroline St. North, Waterloo, ON N2L 2Y5, Canada\label{aff6}
    }

    \date{\today}

    \abstract{
    Under the assumption of statistical isotropy, and in the absence of directional selection effects, a stack of voids is expected to be spherically symmetric, which makes it an excellent object to use for an Alcock–-Paczyński (AP) test. This is commonly done using the void-galaxy cross-correlation function (CCF), which has emerged as a competitive probe, especially in combination with the galaxy-galaxy auto correlation function. Current studies of the AP effect around voids assume that the void centre positions transform under the choice of fiducial cosmology in the same way as galaxy positions. We show that this assumption, though prevalent in the literature, is complicated by the response of void-finding algorithms to shifts in tracer positions. Using stretched simulation boxes to emulate the AP effect, we investigate how the void-galaxy CCF changes under AP, revealing an additional effect imprinted in the CCF that must be accounted for. The effect comes from the response of void finders to the distorted tracer field, reducing the amplitude of the AP signal in the CCF, and thus depends on the specific void finding algorithm used. We present results for four different void finding packages: \revolver, \vide, \voxel, and the spherical void finder in the \pylians library, demonstrating how incorrect treatment of the AP effect results in biases in the recovered parameters for all of them. Finally, we propose a method to alleviate this issue without resorting to complex and finder-specific modelling of the void finder response to AP.
    }


    \maketitle

%#######################################################

\section{Introduction}
\label{sec:introduction}

The large-scale structure of the Universe contains a lot of information about the expansion history as well as the growth of structure within it. The standard way to extract this information is to compute the galaxy-galaxy two-point statistics (either the correlation function or its Fourier equivalent, the power spectrum): the isolated baryon acoustic oscillation (BAO) feature measurable in these statistics can be used as a standard ruler to extract information on the background cosmology \citep[e.g.][]{Alam-DR11&12:2015,Alam-eBOSS:2021}, or the full shape of the two-point clustering can be used to extract perturbation information, including the signatures of galaxy velocities which gives rise to redshift-space distortions \citep[RSD;][]{Kaiser_1987}. Over the last decade cosmic voids, and in particular the use of the void-galaxy cross-correlation function (CCF), have become a powerful complementary probe to these standard analysis techniques. To study the void-galaxy CCF one locates voids---regions of low local galaxy density---within a galaxy sample and cross-correlates the positions of void centres with the galaxy positions. The advantage of the void-galaxy CCF comes from two facts: first, absent any directional distortions or selection effects, for a large enough sample of voids the CCF should be spherically symmetric and is therefore a good candidate to perform the Alcock--Paczyński (AP) test on \citep{Alcock_1979,Lavaux:2012}; second, the dominant source of such directional distortions arises from the coherent outflow velocity from the centres of voids toward over-densities at the boundary, but this can be modelled relatively accurately, thus enabling the AP test. A simple model for the outflow velocity motivated from linear perturbation theory \citep{Hamaus:2014a,Cai:2016a,Nadathur:2019a,Paillas:2021,Massara:2022lng} has been shown to work well enough for the level of statistical precision afforded by data from Stage-III galaxy surveys such as BOSS.

The void-galaxy CCF has been measured in data from the SDSS galaxy surveys \citep{York:2000,Eisenstein:2011,Blanton:2017} by several studies \citep[e.g.,][]{Paz:2013,Hamaus:2016,Hawken:2017,Nadathur:2019c,Achitouv:2019,Hawken:2020,Aubert20a,Woodfinden:2022}, and has been used to place tight constraints on the parameter combination $D\subscr{M}H(z)/c$ via the AP test, where $D\subscr{M}(z)$ is the angular diameter distance and $H(z)$ the Hubble rate at redshift $z$, as well as on the growth rate of structure $f\sigma_8$. Various other observables related to voids have also been considered, like the void size function \citep[e.g.][]{Pisani:2015, Nadathur:2016a, Correa:2018vge, Contarini:2022}, gravitational lensing by voids \citep[e.g.][]{Sanchez:2016, Raghunathan:2020, Bonici:2022}, secondary cosmic microwave background (CMB) anisotropies \citep[e.g.][]{Granett:2008,Nadathur:2016b,Alonso:2018,Kovacs:2019} and void clustering \citep[e.g.][]{Chan:2014,Kitaura:2016b,Zhao:2021}. 

In any void study one first needs to make a choice for what constitutes a void. There is no agreed canonical definition and many different void-finding algorithms exist which give rise to different final populations \citep[see, e.g.,][for a comparison of some algorithms, although not including many new ones]{Colberg:2008qg}. As voids are extended objects and in general not spherical, for the same physical void the location of the void centre may also be defined in multiple ways \citep{Nadathur:2015b}. These are not necessarily problems for observational constraints as long as the void population thus defined corresponds to underdense regions without unwanted selection biases, and the outflow velocity around the chosen void centre location can be accurately modelled. 

Another consideration is whether to identify voids through applying the void-finding algorithm directly on the observed redshift-space galaxy positions, or on a galaxy catalogue with large-scale RSD removed \citep{Nadathur:2019c}. The first option is simpler to implement, but leads to a catalogue with selection effects dependent on the void orientation \citep{Nadathur:2019b,Correa:2022,Euclid:2023eom} and additional imprints of RSD on the void centre positions, requiring additional corrections to the modelling \citep[see ][for an extended discussion]{Nadathur:2020}.

To be able to locate voids in observational data,  we need to assume a fiducial cosmology to first convert observed galaxy redshift and sky angles to distances. If the chosen fiducial cosmology does not match the true cosmology of the Universe, the resulting 3D map of galaxy positions is affected by AP distortions. If they can be correctly modelled, they are a source of information on the background cosmology, via the AP test. For the galaxy-galaxy two-point function, modelling the effect of AP distortions is straightforward: since each galaxy position is determined from its measured redshift in the same way, their positions transform under the same relations and the AP distortion in the correlation function can be modelled as a simple rescaling of the pair separation distances along and perpendicular to the line-of-sight by the so-called AP scaling parameters, which relate $D\subscr{M}$ and $H$ in the true cosmology to their values in the assumed fiducial one.

However, for voids the situation is more complex, since there is no physical object associated with the centre of a void whose redshift is directly measured. Instead the void position is determined from the application of an algorithm on the catalogue of very many galaxies, the positions of each of which already contain AP distortions. To simplify the modelling of the void-galaxy CCF, it has thus far been implicitly \emph{assumed} that the resultant void centre position transforms under the choice of fiducial cosmology in the same way as galaxy positions do, and thus that void-galaxy pair separation distances can be rescaled using the scaling parameters as done for galaxy-galaxy separations. In this paper, we examine the validity of this assumption and show that in general it does not hold, as the two operations of "AP distortion" and void-finding do not commute. The validity of the approximation in theoretical models of the void-galaxy CCF derived under this assumption will depend on the specific void-finding algorithm in question and difference between the true cosmological background and the assumed fiducial model (specifically on the degree of anisotropy introduced by the AP distortion). The aim of this paper is to bring attention to this effect, quantify its size for a few common void-finding algorithms, and propose ways of mitigating this issue. We note that accounting for this effect is only important when the effect of the AP distortion is modelled analytically. It is already appropriately included in simulation-based modelling approaches \citep[e.g.,][]{Paillas:2024,Fraser:2024}.

The layout of this paper is as follows: in Sec.~(\ref{sec:distances}) we give an overview of distances in cosmology including the AP effect and RSD, in Sec.~(\ref{sec:theory}) we review the theory model for the void-galaxy CCF, in Sec.~(\ref{sec:methods}) we outline the methods we used and the different void-finder algorithms, in Sec.~(\ref{sec:results}) we show our results for how the different void-finders respond to AP distortions before concluding in Sec.~(\ref{sec:conclusions}). 

%#######################################################

\section{Distances in cosmology}
\label{sec:distances}

\begin{figure*}
	\centering
	\includegraphics[width=2\columnwidth]{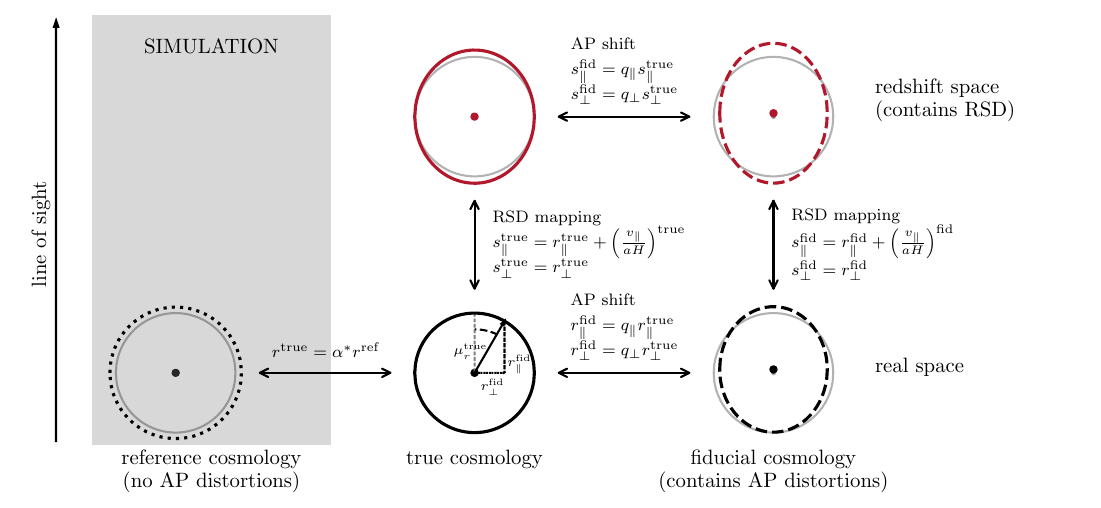}
	\caption{Overview of the different coordinates in use and the transformations between them. Bottom centre is a schematic representation of a void in real space, in the true cosmology. The overall shape of the void is spherical, as expected from statistical isotropy. 
		Going to the top panel (red), we apply a redshift-space distortion, which elongates the void along the line of sight. 
		The right-hand side shows what happens when a wrong cosmology is assumed in the analysis, while the left-hand side shows voids from a simulation used to obtain templates, which are spherically symmetric but their sizes may vary in simulations with different cosmologies. }
	\label{fig:schematic_overview}
\end{figure*}

%---------------------------------------

\subsection{Alcock--Paczyński distortions}
\label{sec:APD}

Instead of physical distances, positions of observed celestial objects are given by angles in the sky, $\theta$ and $\phi$, as well as their measured redshift $z$. The comoving distance to the object is given by
\begin{equation}
	\label{eq:comoving_distance}
	\chi(z) = \int_0^z \frac{c\,{\diff} z}{H(z)} \, .
\end{equation}
As we see, the transformation to physical distances requires us to choose $H(z)$, i.e. assume a fiducial cosmology. If this assumed cosmology is different from the true one, the inferred positions will be wrong, leading to Alcock--Paczyński (AP) distortions \citep{Alcock_1979}. 

For pairs of objects which both have measured redshifts, this logic extends to the distances between them. Let us define the separation vector $\vecr$ of such a pair of objects, which we can decompose along and perpendicular to the line of sight such that
\begin{equation}
	r = \sqrt{r_\parallel^2 + r_\perp^2} \,.
\end{equation}
At redshift $z$, distances perpendicular to the line of sight (LoS) are directly related to the comoving angular diameter distance $D\subscr{M}(z)$, while distances along the LoS are related to the Hubble distance $D\subscr{H}(z) = c/H(z)$. Thus, when the assumed (fiducial) cosmology is slightly different from the truth, these distances are rescaled by the factors
\begin{align}
	\label{eq:AP_apar_dist}
	\apar  \equiv &
	\frac{ D\subscr{H}\superscr{true}(z) }{ D\subscr{H}\superscr{fid}(z) } \,, 
	\\
	\label{eq:AP_aperp_dist}
	\aperp \equiv &
	\frac{ D\subscr{M}\superscr{true}(z) }{ D\subscr{M}\superscr{fid}(z) } \,.
\end{align}
Here, and throughout the paper, we use the superscript $\superscr{true}$ to denote quantities and distances computed in the true cosmology, and superscript $\superscr{fid}$ for quantities and distances in the fiducial one. 

Now we can write the relation between true distances and the ones inferred using the fiducial cosmology as 
\begin{align}
	\label{eq:AP_apar}
	r_\parallel\superscr{true} &= \apar r_\parallel\superscr{fid} \,, 
	\\
	\label{eq:AP_aperp}
	r_\perp\superscr{true}     &= \aperp r_\perp\superscr{fid} \,.
\end{align}
In order to account for the AP effect in modelling the auto-correlation of galaxies (or the power spectrum equivalent in Fourier space), it is therefore sufficient to rescale the separation vectors along and across the line of sight by $\apar$ and $\aperp$, and this approach is extensively used in galaxy survey analyses. The $\apar$ and $\aperp$ parameters can be expressed through an isotropic  volumetric distortion parameter
\begin{equation}
	\label{eq:AP_alpha}
	q = \aperp^{2/3}\apar^{1/3} \,,
\end{equation}
which can be determined through measurement of a feature with an externally-known and well-defined physical scale, such as the BAO peak, and the combination
\begin{equation}
	\label{eq:AP_epsilon}
	\epsilon = \frac{\aperp}{\apar} \,,
\end{equation}
which quantifies the anisotropy introduced due to the AP effect.

Models of the void-galaxy CCF \citep{Paz:2013,Hamaus:2016,Nadathur:2019a,Hamaus:2022} have thus far \emph{assumed} that the distortion of the CCF can be captured using the $(\apar,\aperp)$ or $(q,\epsilon)$ parameter pairs.\footnote{As the absolute void size is taken to be unknown and so, unlike BAO, is not used as a standard ruler, these models are sensitive only to the anisotropy $\epsilon$ and not to $q$.} This is equivalent to assuming that the application of the void-finding algorithm to a set of galaxy positions with or without AP distortions would return exactly the same population of voids, but with the void centre positions shifted by exactly the same amounts as predicted by the AP effect, i.e., \cref{eq:AP_apar,eq:AP_aperp}. Since voids positions are inferred from the large-scale galaxy distribution, rather than being computed from a single measured redshift via the application of \cref{eq:comoving_distance}, this is a non-trivial assumption whose validity remains to be confirmed. In addition to this, there is also the possibility that voids may not form a statistically isotropic population if the chosen fiducial cosmology is far from the truth: that is, void-finding algorithms which have no preferred orientation when applied to a galaxy distribution with no AP distortion, may return samples of voids preferentially aligned either along, or perpendicular to, the LoS when $(\apar,\aperp)\neq(1,1)$.

%---------------------------------------

\subsection{Redshift-space distortions}
\label{sec:RSD}

Peculiar velocities of objects interfere with redshift determination by inducing redshift-space distortions \citep[RSD;][]{Kaiser_1987} along the LoS. Because these velocities contain a wealth of information about the growth of structure, accurately modelling RSD can help us constrain the way cosmic structures evolve over time. 

For two objects with a LoS component of the pairwise peculiar velocity $\mathbf{v}_\parallel$, we can relate the real-space separation $\vecr$ to the distorted separation in redshift space, $\mathbf{s}$, by: 
\begin{align}
	\label{eq:RSD}
	s_\parallel &= r_\parallel + \frac{v_\parallel}{aH} \,, 
	\\ 
	s_\perp &= r_\perp \,, 
\end{align}
where we explicitly see that the distortion only occurs along the LoS, while separation distances perpendicular to the LoS remain unchanged. 

When applying this modelling to the void-galaxy CCF, it is typical to assume that the void positions themselves do not transform under the shift from real to redshift space, so that $v_\parallel$ refers to the peculiar velocities of the galaxies alone. This is sometimes described as an assumption that "void centres do not move". However, as there is no physical object at the void centre whose redshift is measured, the assigned location of the void centre is merely a convention. In practice all studies define the void positions only once, using \emph{either} the observed galaxy positions in redshift space, \emph{or} the galaxy positions with large-scale RSD removed, but not both.\footnote{This choice is made because in general the application of void-finding to the galaxy field with or without RSD does not return the same population of voids \citep{Nadathur:2019b,Correa:2022}, and would therefore violate the assumption of the conservation of the number of void-galaxy pairs under the RSD mapping.} Thus the void positions are invariant under the mapping of galaxies from real to redshift space by construction and any actual motion of void centres is irrelevant---see \citet{Massara:2022lng} for further discussion. 

Note the asymmetry between the conventional treatment of AP and RSD for voids here. The situations are analogous because in both cases void positions are not obtained from an observed redshift of the void, but indirectly determined from the galaxy field---which may or may not contain LoS distortions---and then held fixed and not further transformed. However, while the conventional treatment for RSD (correctly) includes velocity contributions to the inferred positions of galaxies but not to the fixed void positions, the conventional treatment for AP distortions in \cref{eq:AP_apar,eq:AP_aperp} implicitly assumes that the void positions transform in the same way as galaxies.

For simplicity, a schematic overview of the relationship between distances in real and redshift space, as well as the relationship between them in the true and fiducial cosmologies, is given in \cref{fig:schematic_overview}.

%#######################################################

\section{Theory}
\label{sec:theory}

%---------------------------------------

\subsection{AP in the void-galaxy cross-correlation function}
\label{sec:APvoids}

Both of the distortion effects described above leave distinct signatures in the void-galaxy CCF $\xi(\vecr)$, which describes the excess probability over random of finding a galaxy at some distance $\vecr$ from the void. As the signatures that these two effects leave in the CCF are different, we can use it to disentangle and constrain them individually. 

In the literature, the AP treatment for the void-galaxy CCF is based on the one used in galaxy clustering, and uses the equation 
\begin{equation}
	\label{eq:corr_ap}
	\xi\subscr{fid}
	(\vecr\superscr{fid}) = 
	\xi\subscr{true} 
	(\vecr\superscr{true}) \,,
\end{equation} 
which relates the CCF in the true and fiducial cosmologies. \Cref{eq:corr_ap} relies on the implicit assumption that voids in the AP-distorted galaxy field are simply shifted versions of the voids in the undistorted field, and that the shifts in their positions are the same as the shifts that would be incurred by a galaxy located at those original positions. However, the knowledge that recovered void positions depend on the detailed positions of many galaxies, and that they vary strongly with the details of the specific void-finding algorithm used, already suggests that this might not be the case. AP shifts in the tracer field will be translated into shifts in void positions which are not immediately obvious or trivial. 

\begin{figure}
    \centering
    \includegraphics[width=0.9\columnwidth]{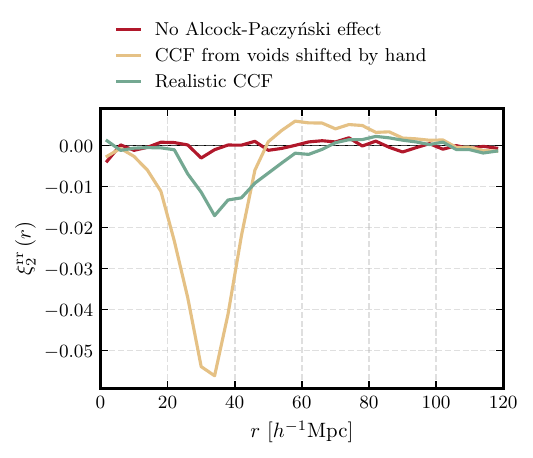}
    \caption{Quadrupole moment of the real-space CCF for voids in the true cosmology (red) and a fiducial cosmology with $\epsilon=1.053$ and $q=1$ (yellow and green). The yellow line is the assumed case where voids move under AP distortion in the same way as halos, while the green line is the more realistic case where voids are found in the distorted halo field. There is a clear additional effect that imprints a signal in $\xir$ when voids are found in a distorted field. For demonstration purposes, the data shown here is obtained using \voxel voids described in \cref{sec:voidfinders}, but similar results apply for all void-finders and can be found in \cref{app:redshiftspace}.}
    \label{fig:demonstration}
\end{figure}

In order to demonstrate this, we construct an idealised test scenario using simulated data (the details of the simulations and void algorithms are described in detail in \cref{sec:methods}). Given a catalogue of simulated (halo) tracers in real space, we shift their positions by stretching their coordinates parallel and perpendicular to an arbitrary LoS direction according to \cref{eq:AP_apar,eq:AP_aperp} in order to mimic an AP distortion, with stretch factors $\apar$ and $\aperp$ chosen for demonstration purposes in order to maintain the volume but introduce a substantial anisotropy, $q=1$ and $\epsilon=1.053$. We create one void catalogue from the application of the void-finding algorithm to these shifted tracers, and a second one from applying the algorithm to the original tracer positions, but then shifting the obtained void positions by $\apar$ and $\aperp$ according to \cref{eq:AP_apar,eq:AP_aperp}. When cross-correlating with the AP-distorted tracer positions, using the first void catalogue provides a realistic representation of the CCF $\xi\superscr{fid}(\vecr\superscr{fid})$ that could be measured in actual survey data, while using the second gives the idealised CCF when void positions are shifted like those of the halo tracers. If the assumption underlying void-galaxy CCF models in the literature is correct, these two approaches should give the same result.

\Cref{fig:demonstration} shows the quadrupole moment of the CCF obtained in these two cases (green and yellow lines respectively). For comparison, we also show the CCF quadrupole for the case with no AP shifts to either tracer or void positions (red). Decomposition of the CCF into Legendre multipoles is standard practice, and the quadrupole moment can effectively summarise the anisotropy of the CCF---since $q=1$ in this demonstration, the monopole of the CCF will be the same in each of the three cases. Unsurprisingly, the quadrupole in the true cosmology (with neither the AP effect nor RSD) is consistent with zero, telling us that the CCF is spherically symmetric, while the other two show significant deviations from spherical symmetry due to the AP effect. It is clear that there is a large difference between the realistic case, where voids were found in the distorted tracer field, and the assumed case where the AP effect affects both tracers and voids in the same way. 

As we will discuss in detail below, this difference in the CCF between the two cases has important consequences for the theory models of the CCF that have been developed so far, leading to systematic modelling errors in scenarios with an anisotropic AP distortion ($\epsilon\neq1$) between the true and fiducial cosmologies. Correcting the models for this requires accounting for the effect of AP shifts on the void-finding algorithm itself. However, since this will be different for different void-finding algorithms, we will forgo trying to analytically describe it and in the following instead attempt to circumvent the problem.

%---------------------------------------

\subsection{RSD modelling}
\label{sec:RSDmodel}

\begin{figure}
    \centering
    \includegraphics[width=0.9\columnwidth]{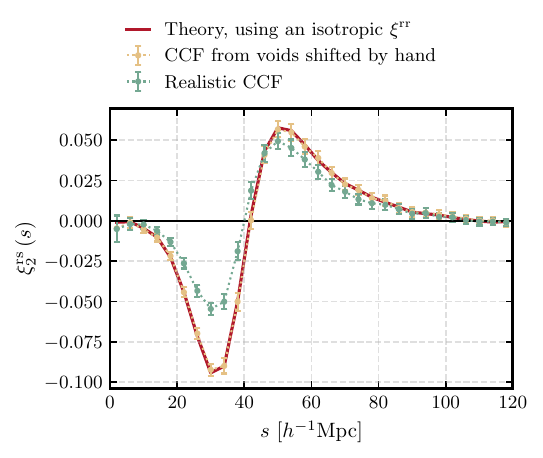}
    \caption{Quadrupole moment of the redshift-space CCF for voids in a fiducial cosmology which differs from the truth ($\epsilon=1.053$ and $q=1$). In red we plot the theory prediction, calculated using a isotropic template $\xir$. This fits the data in yellow, measured from a simulation in which both halos and voids have been moved by hand, while deviating significantly from the more realistic data in green, where voids are found in a distorted halo catalogue. As for \cref{fig:demonstration}, for simplicity only the case of \voxel voids is shown here though similar results apply to all void-finders and can be found in \cref{app:redshiftspace}.}
    \label{fig:demonstration2}
\end{figure}

The second component of models for the CCF between voids and galaxy or halo tracers\footnote{In this paper we will use halo tracers from simulations rather than mock galaxies, but the modelling steps outlined apply to either case.} is accounting for RSD due to the outflow velocities of the tracers around voids. Conservation of void-tracer pairs gives the following relationship:
\begin{equation}
	\label{eq:streaming_model}
	1 + \xi\superscr{s}(\vecs) = 
	\int [1 + \xi\superscr{r}(\vecr)] \, 
	P(v_\parallel,\vecr) \, \diff v_\parallel \,
\end{equation}
between the real-space CCF $\xi\superscr{r}$ and the redshift-space version $\xi\superscr{s}$, describing a convolution with the probability distribution $P(v_\parallel,\vecr)$, describing the distribution of the LoS components of tracer peculiar velocities $v_\parallel$ (as mentioned earlier in \cref{sec:RSD}, this does not include a contribution from the ``motion" of voids). 

Following the notation of \citet{Euclid:2023eom}, we include an extra leading superscript to denote whether the CCF is computed using the positions of voids found in the redshift-space (s) or real-space (r) tracer fields: for example, for the CCF with redshift-space tracer positions, these two scenarios would be denoted by $\xi\superscr{ss}$ and $\xi\superscr{rs}$, respectively, and similarly for the real-space CCF. As has been noted in many works \citep[e.g.,][]{Chuang:2017,Nadathur:2019b,Correa:2022}, performing void-finding on the redshift-space tracer field introduces an additional source of anisotropies, so these CCFs are in general not the same, e.g. $\xi\superscr{ss}\neq\xi\superscr{rs}$. Also, since void numbers are in general not conserved through the application of the void-finding algorithm on different spaces, \cref{eq:streaming_model} can only be used to relate CCFs computed with the same void catalogue on either side of the equation, i.e., $\xi\superscr{ss}$ and $\xi\superscr{sr}$ \emph{or} $\xi\superscr{rs}$ and $\xi\superscr{rr}$, but not $\xi\superscr{ss}$ and $\xi\superscr{rr}$. For the main results in this paper we will follow \citet{Nadathur:2019a,Nadathur:2019c,Woodfinden:2022,Euclid:2023eom} and use $\xi\superscr{rs}$ and $\xi\superscr{rr}$ obtained from voids identified in the real space tracer field without RSD, but in \cref{app:redshiftspace} we demonstrate that the conclusions regarding the importance of the AP effect on void-finding apply to voids in redshift space too.

Assuming spherical symmetry in real space, the peculiar velocity of galaxy or halo tracers around voids can be written as the sum of a coherent radially directed outflow away from the void centre and an additional stochastic component along a random direction, so that the LoS velocity is 
\begin{equation}
	\label{eq:line-of-sight_velocity}
	v_\parallel(r, \mu\subscr{r}) = v_{\rm r}(r) \, \mu\subscr{r} + \tilde{v}_\parallel \,,
\end{equation}
where 
$\mu\subscr{r}=r_\parallel/r$ 
and $\tilde{v}_\parallel$ is the projection of the stochastic component along the LoS. We assume that the LoS velocity distribution $P(v_\parallel,\vecr)$ is a Gaussian with mean $v_{\rm r}\mu_{\rm r}$ and a dispersion which varies with $r$, described in more detail in \cref{sec:velocity} below.

\Cref{eq:streaming_model} makes no reference to the AP effect: it therefore applies only when all quantities on either side of the equation are computed using the same fiducial cosmology and thus with the same AP stretch factors. If this fiducial model does not match the true cosmology, such that $\epsilon\neq1$, the AP effect means that in general $\xir(\vecr)$ is not spherically symmetric (see \cref{fig:demonstration}).\footnote{Note also that if voids are found in redshift space, $\xi\superscr{sr}(\vecr)$ is also not spherically symmetric even in the absence of any AP effect.} However, practical applications of \cref{eq:streaming_model} (or approximations to it, see \citet{Nadathur:2020, Euclid:2023eom} for discussion) in the literature usually provide a spherically symmetric functional form $\xi\superscr{r}(r)$ as a model input on the right hand side. This choice means that what is actually computed on the left-hand side of \cref{eq:streaming_model} is then $\xis\subscr{true}(\vecs\superscr{true})$, i.e. the redshift-space CCF \emph{without} AP effects. To relate this to the final model prediction including both RSD and AP, \cref{eq:corr_ap} is used, but as we have argued, this final association is only valid if void positions truly shift like galaxy positions under the AP transformation.

To demonstrate the issue, in \cref{fig:demonstration2} we repeat the test from \cref{fig:demonstration}, but this time focusing on the quadrupole of the redshift-space CCF, $\xis_2(s)$, again for AP parameters $q=1$ and $\epsilon=1.053$. The green data points are for the realistic CCF measured using voids found in the AP-distorted tracer field, while the yellow data points are for voids found in the original undistorted field and whose positions have subsequently been shifted by hand according to the AP stretch parameters. Only the void catalogue is changed, the positions of the halo tracers including the RSD and AP effects are identical for these two CCFs. The quadrupole moments in the two cases are starkly different.

For comparison, the red curve in \cref{fig:demonstration2} shows the theory model prediction for $\xis_2(s)$, computed using \cref{eq:corr_ap,eq:streaming_model}, assuming the velocity distribution described in \cref{sec:model_inputs}, and assuming a spherically symmetric real-space CCF $\xir(r)$. This assumed isotropic $\xir$ is chosen to match the monopole moment $\xir_0(r)$ of the real-space CCF: since the AP shift in this example is a volume-preserving pure anisotropic distortion with $q=1$, this $\xir_0$ measured with either of the void catalogues is the same and matches the one found with no AP effect at all. The red theory model prediction exactly matches the measured quadrupole for the CCF in the case where void positions have been artificially moved by hand in accordance with the assumed behaviour under the AP effect, but it significantly over-predicts the observed anisotropy in the realistic case. 

To summarise the key result: when voids are found in an anisotropically distorted tracer field, the shift in the void positions obtained compared to those found in the undistorted field is such that it partially cancels out the anisotropy in the CCF. This is an effect that has not been previously noticed, and is not accounted for in current models of the void-galaxy CCF. The model failure occurs only for $\epsilon\neq1$, and increases with $|\epsilon-1|$: therefore it is not seen when the fiducial cosmology is chosen to be close to or exactly match the true cosmology, as is often the case in tests of systematic errors on mocks \citep[although see][as examples of studies that examine systematic errors associated with the choice of fiducial cosmology]{Nadathur:2020,Aubert20a,Woodfinden:2022}. If not correctly accounted for, this can possibly bias the central value of the posterior for $\epsilon$ if the fiducial cosmology lies a long way from the truth. Perhaps more worrying is that this has a strong effect on derived errors as we will see below. First, we describe in more detail how model predictions for the CCF can be adjusted to mitigate this effect.

%---------------------------------------

\subsection{Model inputs}
\label{sec:model_inputs}

A full evaluation of \cref{eq:streaming_model} requires the specification of the real-space CCF $\xir$ and the velocity distribution $P(v_\parallel,\vecr)$. We now describe the details of how these are computed for the calculations performed in this paper.

\subsubsection{Velocity distribution}
\label{sec:velocity}

We model the velocity distribution $P(v_\parallel,\vecr)$ as an isotropic Gaussian distribution dependent on the distance from the void centre:
\begin{equation}
	\label{eq:velocity_PDF}
	P(v_\parallel,r) = 
	\frac{1}{ \sqrt{2\, \pi\, \sigma_{v_\parallel}^2(r)} } 
	\exp\left\{
		-\frac{[v_\parallel-\mu_r v_{\rm r}(r)]^2}{ 2\, \sigma_{v_\parallel}^2(r)} 
	\right\} \,.
\end{equation}
This distribution depends on two functions to be specified: the coherent radial outflow velocity $v_{\rm r}(r)$ and the velocity dispersion $\sigma_{v_\parallel}(r)$.

The outflow velocity $v_{\rm r}(r)$ is usually assumed to be related to the enclosed matter density profile $\Delta(r)$ for voids via a linearised form of the continuity equation, $v_{\rm r}(r) = -\frac{1}{3} f\,a\,H\,r\,\Delta(r)$, where $f$ and $a$ are the growth rate and scale factor, respectively. This approximation works reasonably well \citep[e.g.][]{Hamaus:2014fma,Nadathur:2019a}, although \citet{Massara:2022lng} highlighted shortcomings in some circumstances. The matter profile $\Delta(r)$ is in turn either determined from the monopole moment $\xir_0(r)$ of the real-space CCF via an assumed effective linear bias relationship \citep[e.g.,][]{Aubert20a,Hamaus:2022,Schuster:2023}, or taken from a template function $\Delta\superscr{ref}$ calibrated from simulations \citep[e.g.,][]{Nadathur:2019c,Nadathur:2020,Woodfinden:2022}. In the latter approach, as the template $\Delta\superscr{ref}(r)$ is empirically found to scale proportionally with the amplitude of matter density perturbations, such that $\Delta\superscr{true}(r)=\left(\sigma_8\superscr{true}/\sigma_8\superscr{ref}\right)\Delta\superscr{ref}$ \citep{Nadathur:2019c}, where the superscripts true and ref refer to the true cosmological model and that of the simulation from which the template was calibrated. Together with the continuity equation, this empirical scaling results in the relationship
\begin{equation}
    \label{eq:vr_scaling}
    \left(\frac{v_{\rm r}}{aH}\right)\superscr{true} = \frac{(f\sigma_8)\superscr{true}}{(f\sigma_8)\superscr{ref}}\left(\frac{v_{\rm r}}{aH}\right)\superscr{ref}\,,
\end{equation}
i.e. the velocity profile inherits a dependence $\propto f\sigma_8$. Note that the $aH$ factors on either side of the equation also allow for minor differences between the true effective redshift of the data and the simulation redshift used for template calibration to be included. 

In this paper, we adopt a modification of the template calibration: instead of calibrating a template $\Delta\superscr{ref}(r)$ from simulations, we directly measure a velocity profile template $v\subscr{r}\superscr{ref}(r)$. This means we can isolate imperfections in the CCF model that arise from the AP distortions of interest here from those that may be related to residual deviations from the assumption of purely linear dynamics relating $v_{\rm r}$ and $\Delta$. Nevertheless, we continue to assume that the scaling of $v_{\rm r}$ with the growth factor $f\sigma_8$ implied by \cref{eq:vr_scaling} holds.

In the same way, we also calibrate a template velocity dispersion function $\sigma\superscr{ref}_{v_\parallel}(r)$, and take the true velocity dispersion to be proportional to this template, with a free amplitude that is treated as a nuisance parameter and marginalised over when reporting cosmological constraints \citep[see ][for details]{Euclid:2023eom}. 

In practice, the template functions are calibrated for voids with an average size defined by some chosen cuts on the void sample. The apparent size of voids in the real Universe, where $q$ is unknown, may differ from the physical size of voids in the simulation, where $q=1$, so the same apparent size cuts may result in a void sample of different average physical size in the two situations. To account for this, we allow the model functions in the true cosmology to differ from the reference templates through an isotropic rescaling of all distance scales, setting $X\subscr{true}(r\superscr{true})=X\subscr{ref}(\astar r\superscr{true})$, where $X$ represents the model function in question and $\astar$ is a free parameter. 

If changes from the template occur purely due to the AP volume dilation, we should have $\astar=q=\aperp^{2/3}\apar^{1/3}$ and the effect of such rescaling on the model in \cref{eq:streaming_model} is exactly cancelled by the opposite shift when the AP model is applied to the redshift space CCF in the final step using \cref{eq:corr_ap}. This ensures that the observed void size cannot be used as a standard ruler and no constraints on $q$ are obtained. The same cancellation also automatically occurs in alternative approaches \citep[e.g.][]{Hamaus:2022} in which distances from the void centre are always scaled in units of apparent void size. To allow for more general deviations in shape from the calibrated templates, we can also allow $\astar$ to differ from $q$, but in this case the two parameters remain very degenerate and so it is still not possible to constrain either of them.

\subsubsection{Real-space CCF}
\label{sec:realspaceCCF}

The predictions obtained from \cref{eq:streaming_model} and models related to it are most sensitive to the form of the real-space CCF $\xir(\vecr)$, so this is the most important input to get right. Various approaches have been used to determine $\xir(\vecr)$, including using an analytic formula \citep{Hamaus:2014a}, using a template calibrated from simulations \citep{Nadathur:2019c,Nadathur:2020,Woodfinden:2022}, and determining it directly from data \citep{Hamaus:2022,Euclid:2023eom}. We will consider examples of the last two of these methods in the following.

The first, template-based, method involves using simulated mock data that is designed to approximately match the tracer number density and large-scale clustering amplitude of the real data. Through applying the void-finding and CCF measurement steps on such a mock, we can determine a reference template function $\xir\subscr{ref}$. As the effects of RSD and AP distortions can be exactly removed in the mock where tracer velocities and the true cosmology are known, this template is usually perfectly isotropic to within measurement errors, $\xir\subscr{ref}=\xir\subscr{ref}(r)$. As for the velocity template functions above, we allow an arbitrary isotropic rescaling of distance scales, $\xir\subscr{true}(r\superscr{true})=\xir\subscr{ref}(\astar r\superscr{ref})$, which ensures that the template function is not used as a cosmological standard ruler.

As the reference template function is isotropic, the template-based model relies on the assumption that the AP effect on void-finding does not introduce anisotropies in the true real-space CCF. The model also implicitly assumes that AP distortions can be modelled as though void centres move as galaxy or halo tracers. We have already shown that these two assumptions do not hold: the performance of the template-based model then allows us to quantify the effect these assumptions have on the final model performance.

The second approach involves determining the real-space CCF $\xir(\vecr)$ from the data. \citet{Hamaus:2022} do this by using the redshift-space data to measure the projected CCF $\xi_\mathrm{p}(s_\perp)$, which is independent of RSD effects, and using a deprojection technique \citep{Pisani:2014} to obtain the real-space CCF. The deprojection step assumes spherical symmetry of the real-space CCF, therefore suffers from the same limitations with respect to the AP effect as the template method. \citet{Euclid:2023eom} propose a different method, which is based upon a reconstruction technique to approximately remove RSD from the galaxy catalogue \emph{before} void-finding; using this reconstructed catalogue we can measure $\xir\subscr{fid}(\vecr)$ in the fiducial cosmology. \citet{Euclid:2023eom} demonstrated that this method can remove the effects of RSD to recover the real-space CCF sufficiently well for modelling purposes. We will assume that this RSD removal can be performed perfectly, with no residuals. This is not a fully realistic assumption in practice, but it helps us to separate out the additional AP effect we wish to determine here from any possible confounding RSD residuals. We will therefore use the known tracer velocities in our simulations to exactly subtract RSD in determining $\xir\subscr{fid}(\vecr)$, rather than using approximate reconstruction methods.

From the measured $\xir\subscr{fid}(\vecr)$ and for given AP scaling parameters $\aperp$ and $\apar$, we infer the true real-space CCF using
\begin{equation}
    \label{eq:RSD_true}
	\xi\superscr{rr}\subscr{true}
		(r_{||}\superscr{true}, r_{\perp}\superscr{true}) =
	\xi\superscr{rr}\subscr{fid}
		(\apar r_{||}\superscr{true}, \aperp r_{\perp}\superscr{true})\,. 
\end{equation}
Importantly, if void-finding in the AP-distorted galaxy or halo field introduces additional anisotropies beyond those included in the simple model where void positions transform in the same way as tracers, as indicated by \cref{fig:demonstration}, these are automatically included in the measured $\xir\subscr{fid}(\vecr)$ via \cref{eq:RSD_true}. That is, even when the correct $\aperp$ and $\apar$ parameters corresponding to the true cosmology are used in \cref{eq:RSD_true}, the resulting $\xi\superscr{rr}\subscr{true}$ function that is obtained \emph{need not be spherically symmetric}. Anisotropies due to void-finding are thus ``carried through" the model computation in \cref{eq:streaming_model}. 

\begin{figure}
    \centering
    \includegraphics[width=\columnwidth]{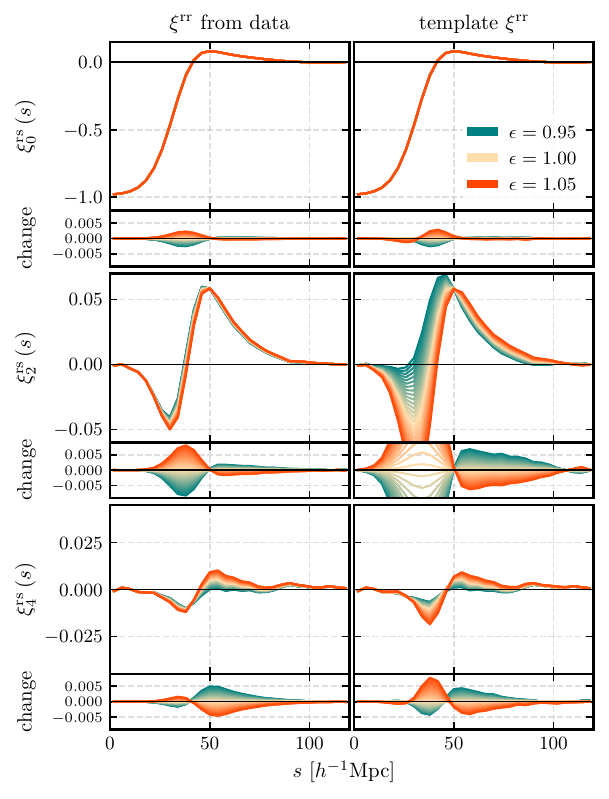}
    \caption{Theoretical redshift-space CCF calculated for a range of fiducial cosmologies. The left-hand side shows how the theory responds to changes in the AP distortion parameter $\epsilon$ when the real-space CCF $\xir$ comes from the data, while the right-hand side shows the same for the case when $\xir$ is given as a template from simulations. From top to bottom, the three large panels show the monopole, quadrupole and hexadecapole, while the residuals show the difference between the multipoles at a given $\epsilon$ and the multipoles when $\epsilon=1$. }
    \label{fig:theory_with_epsilon}
\end{figure}

In \cref{fig:theory_with_epsilon} we compare the model predictions using the possibly anisotropic $\xir$ determined from data and those obtained when assuming an isotropic $\xir$ taken from a template in a fixed cosmology. The results show that assuming isotropy of $\xir$ predicts a much greater model sensitivity to the AP parameter ratio $\epsilon=\aperp/\apar$, particularly in the quadrupole moment. This is because this approach neglects the anisotropic selection effect of void-finding performed in a distorted tracer field, which tends to partially cancel the anisotropy introduced in the CCF. We show below that accounting for the anisotropic effect in void-finding results in much more accurate model predictions when $\epsilon\neq1$.

%#######################################################

\section{Methods}
\label{sec:methods}

%---------------------------------------

\subsection{Simulations}
\label{sec:simulations}

\begin{table}[]
	\setlength\extrarowheight{3pt}
	\caption{Combinations of Alcock--Paczyński scaling parameters used in the analysis. We differentiate between two groups of cases: flat $\Lambda$CDM fiducial cosmologies with different values of $\Omega\subscr{m}$, and a more general fiducial cosmology where $q=1$ but $\epsilon$ varies (isovolumetric distortion). }
	\label{tab:all_cases}
	
	\centering
	\begin{tabular*}{0.95\linewidth}{@{\hspace{5pt}} rc @{\extracolsep{\fill}}cccc@ {\hspace{20pt}} }
		\hline\hline
		& $\epsilon$ & $q$ & $\apar$ & $\aperp$ &  $\Omega\subscr{m}$ \\
		\hline
		& 0.978 & 1.031 & 1.046 & 1.023 & 0.388 \\
		& 0.984 & 1.022 & 1.033 & 1.017 & 0.367 \\
		& 0.990 & 1.013 & 1.020 & 1.010 & 0.347 \\
		& 0.997 & 1.005 & 1.007 & 1.003 & 0.327 \\
		$\triangleright$ & 1.000 & 1.000 & 1.000 & 1.000 & 0.317 \\
		& 1.003 & 0.995 & 0.993 & 0.997 & 0.308 \\
		& 1.010 & 0.986 & 0.979 & 0.990 & 0.287 \\
		& 1.018 & 0.977 & 0.966 & 0.983 & 0.267 \\
		& 1.025 & 0.967 & 0.951 & 0.976 & 0.248 
		\vspace{5pt}\\
		& 0.952 & 1.000 & 1.033 & 0.984 & $\ldots$ \\
		& 0.978 & 1.000 & 1.015 & 0.993 & $\ldots$ \\
		& 0.984 & 1.000 & 1.011 & 0.995 & $\ldots$ \\
		& 0.990 & 1.000 & 1.007 & 0.997 & $\ldots$ \\
		& 0.997 & 1.000 & 1.002 & 0.999 & $\ldots$ \\
		$\triangleright$ & 1.000 & 1.000 & 1.000 & 1.000 & 0.317 \\
		& 1.003 & 1.000 & 0.998 & 1.001 & $\ldots$ \\
		& 1.010 & 1.000 & 0.993 & 1.003 & $\ldots$ \\
		& 1.018 & 1.000 & 0.988 & 1.006 & $\ldots$ \\
		& 1.025 & 1.000 & 0.983 & 1.008 & $\ldots$ \\
		& 1.053 & 1.000 & 0.966 & 1.017 & $\ldots$ \\
		\hline
	\end{tabular*}
\end{table}

For the analysis presented, we made use of 200 Quijote simulations\footnote{\url{https://quijote-simulations.readthedocs.io/en/latest/}} \citep{Villaescusa-Navarro:2019bje} at redshift $z=0.5$, with a box size of side $L=1000\,\hMpc$. All of the boxes are at the fiducial Quijote cosmology, a flat $\Lambda$CDM cosmology with $\Omega\subscr{m}=0.3175$, $\Omega\subscr{b}=0.049$, $h=0.6711$, $n\subscr{s}=0.9624$, $\sigma\subscr{8}=0.834$. Despite our theory describing the void-galaxy CCF, we make use of the provided halo catalogues, with friends-of-friends (FoF) halos found with a halo finder implemented in \pylians.\footnote{\url{https://pylians3.readthedocs.io/en/master/index.html}} This is equivalent to assuming every halo has one central galaxy, resembling a typical sample of luminous red galaxies (LRGs), which have a low fraction of satellite galaxies. 

Each halo catalogue contains positions and velocities for about $3.09\pow{5}$ halos in real space, corresponding to a number density of $3.09\pow{-4} \,h^3{\rm Mpc}^{-3}$. We used the provided halo velocities to construct corresponding redshift-space halo catalogues using \cref{eq:RSD} and assuming a line of sight along the direction of the $z$-axis of the box.  

To avoid converting Cartesian positions of halos into angles and redshift, and re-converting them back into Cartesian coordinates while assuming a different cosmology, we instead stretched the simulation boxes directly. To do this, we applied the AP scaling factors $\apar$ and $\aperp$ to the halo positions in both real and redshift space, keeping the $z$-axis as the LoS direction and using \cref{eq:AP_apar,eq:AP_aperp}. In the case where the scaling factors are greater than $1$, we first duplicated the boxes along all the axes to make sure the  entire box volume remained filled after compression. Finally, we cut the box down to its original size of $1000\,\hMpc$. 

This process results in cubic boxes which no longer satisfy periodic boundary conditions. While we can mostly ignore this for the purposes of void finding, we in general do not trust the results near the edges of the box. Therefore, for the purpose of measuring the CCF and velocity profiles, we conservatively cut $150\,\hMpc$ on all sides of the box, which is a scale greater than the largest void-galaxy pair separations considered. This leaves us with a volume that is about three times smaller than the original box volume. 

We consider two groups of cases for the choice of $\apar$ and $\aperp$: flat $\Lambda$CDM cosmologies with different values of $\Omega\subscr{m}$ and cosmologies corresponding to isovolumetric distortion ($q=1$ and $\epsilon\neq 1$). All combinations of $\apar$ and $\aperp$ (and $\epsilon$ and $q$) used in this paper are given in \cref{tab:all_cases}. 

%---------------------------------------

\subsection{Void finding}
\label{sec:voidfinders}

\begin{figure}
	\centering
	\includegraphics[width=0.9\columnwidth]{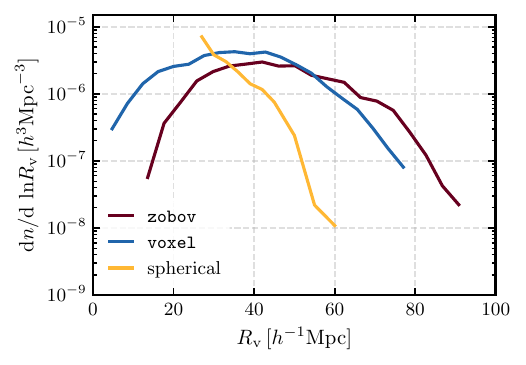}
	\caption{Void size function (VSF) for different void finders. We plot only one VSF for the two \zobov-based finders since the only difference between them is the void centre definition, not the number of voids or their sizes. }
	\label{fig:void_size_function}
\end{figure}

The process of void finding can be conceptualized as a non-linear transformation of the tracer field, leading to the identification of underdense regions. While different void-finding algorithms all aim to recover these underdense regions, they all employ different methodologies, resulting in vastly different voids, with different positions and statistical properties. In this paper, we have focused on four void-finders, which use three different algorithms, and explored how the AP effect influences the properties and characteristics of the voids identified by these algorithms. 

The first two algorithms, \zobov \citep{Neyrinck:2007gy} and \voxel \citep{Revolver}, use a watershed-based technique, while the spherical void finder is based on the excursion set formalism \citep{Banerjee16}. Here we will present a short overview of all of these algorithms as they are used on cubic simulation boxes.

%--------
\subsubsection{\zobov}

\zobov \citep{Neyrinck:2007gy} is a watershed void-finder that uses a Voronoi tessellation approach to estimate the density field. In this approach the cosmological volume is divided such that each tracer particle is assigned a Voronoi cell, an area of space which is closer to that specific particle than it is to any other. These cells vary in size, with larger cells corresponding to regions where tracer particles are further away from each other, and smaller cells indicating areas of higher density (or closer tracer proximity). 

The volume of the Voronoi cells helps us estimate the density field ($\rho \propto 1/V$), as the largest cells pinpoint the locations of local minima, which serve as the starting points for the watershed algorithm. Additional neighbouring cells are progressively included in the void's volume, with their densities increasing until a cell with lower density (a saddle point) is encountered. As this approach results in irregular void volumes, it is helpful to define an effective void radius $R\subscr{v}$ as the radius of a sphere with the same volume as the void: 
\begin{equation}
	R\subscr{v} = \left({\; \frac{3}{4\,\pi} \sum_i V_i\;} \right)^\frac{1}{3}\,,
\end{equation}
where $V_i$ is the volume of the $i$-th Voronoi cell, and the sum goes over all the cells belonging to the void. From here we can choose one of two different ways of estimating the void centre: circumcentre -- used in the \revolver\footnote{\url{https://github.com/seshnadathur/Revolver}} \citep{Revolver} void-finding package -- and barycentre -- used in the \vide\footnote{\url{https://bitbucket.org/cosmicvoids/vide_public/src/master/}} \citep{Sutter_2015} void finder. 
 
In the \revolver circumcentre definition, we find the four most underdense mutually adjacent Voronoi cells belonging to the void. The tracer particles belonging to them form a tetrahedron, the circumcentre of which is defined as the centre of the void. Voids defined in this way naturally tend to be particularly underdense in the centre, going down to $\delta=-1$, and have a characteristic knee in the density profile, corresponding to the four tracer particles used to define the centre. 

The Void IDentification and Examination toolkit \citep[\vide;][]{Sutter_2015} defines the location of the void centre as the volume-weighted barycentre of all the tracer particles belonging to the void. This gives us a non-local definition of the void centre and thus, while the voids identified by \revolver and \vide are identical, the properties of the CCF measured using void centre positions differ significantly.

Since void finding on a discrete tracer field can result in many small, spurious voids, we impose a selection criterion on our voids, removing all voids with a radius $R\subscr{v}$ smaller than $30\,\hMpc$. This also removes true small voids which are mostly governed by non-linear evolution and whose properties are not accurately described by the model given in \cref{sec:theory}.

%--------
\subsubsection{\voxel}

Unlike \zobov, \voxel is a void-finding algorithm that uses a particle-mesh interpolation technique to accurately estimate the density field. In this approach, tracers are placed on a three-dimensional grid of cells (voxels), with the grid size determined by the average number density of tracers $\bar n$, such that the side-length of each voxel is $a\subscr{vox}=0.5\left(4\pi\bar n/3\right)^{-1/3}$, and the density is proportional to the number of tracers in the voxel. This interpolation scheme is easily parallelizable and faster than other density estimation methods, such as the one used in \zobov, scaling simply as the number of tracers, $\propto N$. 

Following the estimation of the density field, a Gaussian filter with a smoothing length $r\subscr{s}=\bar{n}^{-1/3}$ is applied and voids are identified as local minima within this smoothed density distribution. To determine the size and non-overlapping boundaries of each void, a watershed algorithm progressively includes neighbouring voxels until a critical point is reached where the density of the next cell is lower than that of the previous one. The centre of a \voxel void is naturally defined as the position of the lowest density voxel belonging to it. As this is a locally defined centre, it is very sensitive to small scale fluctuations of the density field; however, this does not seem to significantly affect the statistical properties of \voxel voids. 

We once again find irregularly shaped voids and assign each an effective radius
\begin{equation}
	R\subscr{v} = \left({\; \frac{3}{4\,\pi} N\subscr{vox}V\subscr{vox}\;} \right)^\frac{1}{3}\,,
\end{equation}
where $N\subscr{vox}$ is the number of voxels belonging to the the void, and $V\subscr{vox}$ is the voxel volume. Just like for \zobov, we impose a radius cut on the \voxel void catalogue. To keep the comparison easier, we use the same criterion of $30\,\hMpc$ and remove any voids with a radius smaller than that. 

Though in this work we were limited to cubic boxes, in \citet[in prep.]{Fraser:2024} the \voxel algorithm was updated to handle non-cubic periodic boundary conditions. This facilitates the identification of voids in the distorted boxes without the need for the trimming procedure described in \cref{sec:simulations}. 

%--------
\subsubsection{Spherical void finder}

The spherical void finder used in this paper is implemented in the \pylians library,\footnote{\url{https://github.com/franciscovillaescusa/Pylians3}} and identifies non-overlapping spherical underdensities. As input, it requires a list of radii and an enclosed-overdensity threshold $\Delta\subscr{v}$ which will be used to define a void. For the analysis presented here, we used a threshold of $\Delta\subscr{v} = -0.7$. 

The void finder starts by placing tracer particles on a grid, and then smoothing the field with a top-hat filter of radius $R\subscr{v}$, where $R\subscr{v}$ is the largest void radius in the input list. It then identifies all regions where the smoothed enclosed overdensity is below the threshold. Going from the most underdense of these regions, we define  a region as a void if it does not overlap with an already defined void. Though these voids can in principle have radii larger than the top-hat smoothing scale, they are all assigned the same radius $R\subscr{v}$. The smoothing and identification process is then repeated for all the given void radii, from largest to the smallest, until we have a list of non-overlapping voids which fill the entire cosmological volume. 

The voids we obtain in this way are effectively binned in radius, with the bins defined by the input list. For the analysis presented here, we used only the radius bin $27\leq R\subscr{v}<30 \,\hMpc$ to calculate the CCF and the velocity template functions. It is useful to note that, while void radii of the various watershed algorithms can be roughly compared, radii of spherical voids have a different definition and therefore cannot be trivially compared to the watershed ones. 

%---------------------------------------

\subsection{Measuring velocity templates}
\label{sec:templates}

\begin{figure}
	\centering
	\includegraphics[width=0.9\columnwidth]{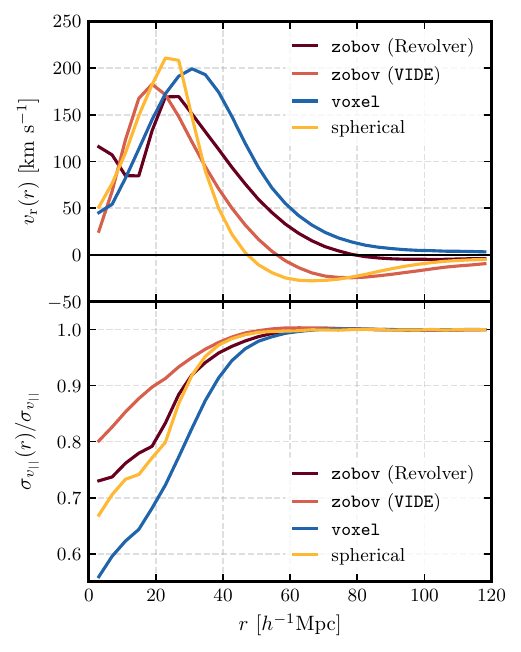}
	\caption{Radial velocity and normalized line-of-sight velocity dispersion templates for different void finders. The templates are taken at a reference cosmology which matches the truth, presenting an idealized scenario. }
	\label{fig:templates}
\end{figure}

\begin{figure*}
	\centering
	\includegraphics[width=1.8\columnwidth]{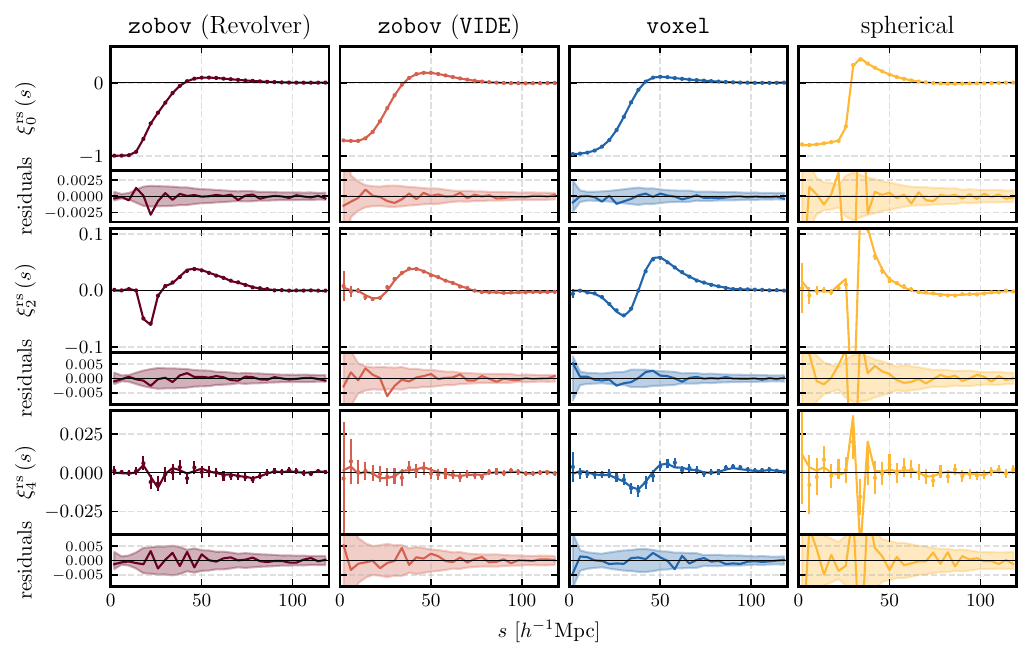}
	\caption{Data and the model for all void finders considered in the paper, calculated in the true cosmology. We see the differences in shapes of various voids, such as the characteristic knee in the monopole of \zobov voids from \revolver, and the sharp features of spherical voids. }
	\label{fig:true_fit}
\end{figure*}

Using the real-space positions of voids and halos in the true cosmology, and with the available halo peculiar velocities, we measured the radial outflow velocity and the LoS velocity dispersion. In practice, these templates would be measured from a simulation that is at some reference cosmology and does not contain AP distortion. However, since we instead measure them at the true cosmology, we construct an idealized scenario where the reference cosmology happens to be matching the true one (meaning that the simulation perfectly describes the Universe). We do this to demonstrate that the effect of AP distortions on void finding is there even in an ideal scenario, where we have perfect templates that match the truth.

To measure radial velocities around voids, the voids are stacked and the radial component of the halo velocities around them are binned in spherical shells, while averaging over all the void-halo pairs in the bin regardless of angle $\mu$. This gives slightly different velocity profiles $v_{\rm r}\superscr{ref}(r)$ for each void finder, as seen in the top panel of \cref{fig:templates}. Regardless of their differences, the main feature is present in all: a peak around the steep edge of the void, which indicates a coherent radial outflow form the centre of the void. The reference velocity templates for each void type are rescaled according to \cref{eq:vr_scaling} to obtain the model $v_{\rm r}(r)$ used in \cref{eq:velocity_PDF} for the velocity probability distribution function (PDF).

To obtain the dispersion profile, we compute the modulus of the halo velocity in each of the bins, $\langle|\mathbf{v}|^2\rangle$, and thus measure \citep[see ][]{Fiorini:2022srj}
\begin{equation}
	\label{eq:sigmav}
	\sigma_{v_\parallel}(r) = \sqrt{ \frac{\langle|\mathbf{v}|^2\rangle - v\subscr{r}^2}{3} }\,,
\end{equation}
for each void sample, which are shown in the bottom panel of \cref{fig:templates} (normalised by the dispersion $\sigma_v\equiv\sigma_{v_\parallel}(r\to \infty)$ at large scales). We define the reference templates $\sigma\superscr{ref}_{v_\parallel}(r)$ from these measured functions, leaving the amplitude $\sigma_v$ as a free parameter. 

%---------------------------------------

\subsection{Measuring the void-halo CCF}
\label{sec:ccf}

After again stacking all voids that survived the radius cuts, as well as removing any voids near the edges of the simulation box (see \cref{sec:simulations}), we used a modified and Python-wrapped version of CUTE\footnote{\url{https://github.com/seshnadathur/pyCUTE}, based on \url{ https://github.com/damonge/CUTE}} \citep[Correlation Utilities and Two point Estimation;][]{2012arXiv1210.1833A} to count void-halo pairs $D\subscr{v}D\subscr{g}$. It estimates the CCF via:
\begin{equation}
	\label{eq:estimator}
	\xi(r, \mu)=\frac{V}{ N\subscr{v} N\subscr{g} }
	\frac{D\subscr{v} D\subscr{g}(r,\mu)}{ v(r,\mu) } \,,
\end{equation}
where $V$ is the box volume, $N\subscr{v}$ and $N\subscr{g}$ are the numbers of voids and galaxies, respectively, and $v=4\pi r^2\Delta r \Delta\mu$ is the volume of the $\Delta r \Delta\mu$ bin. We used 100 angular bins in range $\mu\in(0,1)$, and 30 radial bins up to $r=120 \hMpc$. Correlating the voids with real-space halos gives us $\xir$, while correlating the same voids with redshift-space halos gives us $\xis$. This process is applied to all 200 Quijote mocks, and the resulting CCFs are averaged over these 200 mock realisations.

The CCF measurement for both real and redshift space is done for all of the fiducial cosmologies listed in \cref{tab:all_cases}, with the one done at the true cosmology (no AP effect) serving also as a template $\xi\subscr{rr}\superscr{ref}$. As with the velocity, these templates are then at the true cosmology and present an idealized scenario. 

%---------------------------------------

\subsection{Covariance matrix}
\label{sec:covmat}

\begin{figure}
	\centering
	\includegraphics[width=0.9\columnwidth]{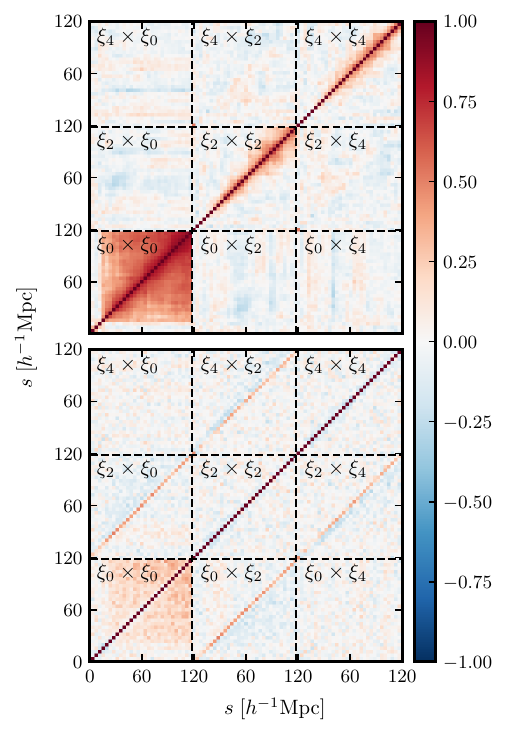}
	\caption{Normalized data-only (top) and full (bottom) covariance matrix. We see that the off-diagonal correlation structure is much less pronounced in the full covariance, compared to the data-only one. }
	\label{fig:correlation_matrix}
\end{figure}

Since we consider two distinct cases, $\xir$ from data or as a template, we construct two covariance matrices. When $\xir$ is supplied as a template, the covariance matrix is the standard, data-only covariance. It is constructed from $N\subscr{m}=200$ mocks, using the equation
\begin{equation}
	\label{eq:covariance}
	\mathsf{C}_{ij}\superscr{data} =
	\frac{1}{N\subscr{m} - 1} \sum_{k=1}^{N\subscr{m}} 
	\left( \mathbf{\xis}_i^{(k)} - \langle \mathbf{\xis}_i \rangle \right)
	\left( \mathbf{\xis}_j^{(k)} - \langle \mathbf{\xis}_j \rangle \right) 
	\, ,
\end{equation}
where $\mathbf{\xi}_{i}^{(k)}$ is the $i\superscr{th}$ bin of the redshift-space CCF for the $k\superscr{th}$ mock, while $\langle \mathbf{\xis} \rangle$ is the mean over all $N\subscr{m}$ mocks. A covariance matrix constructed like this represents the error budget for a volume of $(700\hMpc)^3$, which is the volume of one mock after we cut out the edges (see \cref{sec:simulations}). For an easier demonstration of the issues we present in this paper, we have chosen to rescale the covariance to a more realistic survey volume. We chose a DESI-like survey \citep{DESI:2016fyo}, with a volume about 25 times that of our cut simulation box\footnote{For DESI luminous red galaxies (LRGs). }, which means we divide the covariance in \cref{eq:covariance} by 25, assuming a linear scaling with volume. 

When the real-space CCF $\xir$ is taken from the data itself, the theory calculated this way is naturally correlated with the data. This correlation needs to be accounted for and added into the covariance. We can write \citep{Euclid:2023eom}:
\begin{align}
	\label{eq:DMTcov}
	\mathsf{C} ({\rm data} - {\rm theory}) = 
	\mathsf{C} ({\rm data}) 
	&+ \mathsf{C}({\rm theory}) \nonumber \\ 
	&-2\mathsf{C}({\rm data, theory}) \,,
\end{align}
where the first term is the one given by \cref{eq:covariance}, the second is calculated in a similar way but with the theoretical $\xis$ instead of the measured one, while the last term captures the cross-covariance between the two. This cross-covariance term brings down the total variance and, therefore, results in tighter error bars. 

It would be logical to conclude that using the full covariance $\mathsf{C}\superscr{tot}$ would then result in tighter constraints on $\epsilon$, but one look at \cref{fig:theory_with_epsilon} shows why such a conclusion might be wrong. Namely, our model responds more strongly to changes in $\epsilon$ when we supply a template real-space CCF, resulting in tighter constraints (see \cref{sec:results}). 

Covariances estimated like this, using a finite number of mocks, are inherently uncertain. Therefore, we have to propagate this uncertainty to the likelihood, which we do using the prescription described in \citet{Percival:2021cuq}. We write the likelihood as 
\begin{equation}
	\label{eq:likelihood}
	\log \mathcal{L} = 
	-\frac{m}{2} \log\left( 1 + \frac{\chi^2}{N\subscr{m} - 1} \right) \,,
\end{equation}
with $\chi^2$ given by
\begin{equation}
	\label{eq:chisq}
	\chi^2 = 
	\left( \xis\subscr{\,theory} - \xis\subscr{\,data} \right)
	\mathsf{C}^{-1}
	\left( \xis\subscr{\,theory} - \xis\subscr{\,data} \right)
	\superscr{T} \,. 
\end{equation}
The power-law index $m$ is defined as:
\begin{align}
	\label{eq:likelihood_indices}
	m &= n\subscr{\theta} + 2 + \frac{N\subscr{m} - 1 + B (n\subscr{d} - n\subscr{\theta})}{1 + B (n\subscr{d} - n\subscr{\theta})} \,, \\
	B &= \frac{N\subscr{m} - n\subscr{d} - 2}{(N\subscr{m} - n\subscr{d} - 1) (N\subscr{m} - n\subscr{d} - 4)} \,,
\end{align}
where the $N\subscr{m}=200$ is, as before, the number of mocks used, and $n\subscr{d}=90$ is the number of data points (30 radial bins for 3 multipoles). Finally, $n\subscr{\theta}$ is the number of independent parameters. Though our parameter space is given by five parameters, $\{ f\sigma_8, \sigma_v, \epsilon, q, \astar \}$, $q$ and $\astar$ are completely degenerate, which leaves us with $n\subscr{\theta}=4$.

To minimize the likelihood given in \cref{eq:likelihood}, we sample the parameter space using the Monte Carlo Markov Chain (MCMC) sampler in \texttt{Cobaya}\footnote{\url{https://cobaya.readthedocs.io/en/latest/}} \citep{Torrado:2020dgo}, with the model and likelihood implemented in \texttt{victor}\footnote{\url{https://github.com/seshnadathur/victor}}. Convergence of the chains was ensured using the Gelman-Rubin $R-1$ statistic, using the convergence criterion $R-1<0.01$, and 20\% of the initial steps were discarded as burn-in.

%#######################################################

\section{Response of void finders to AP distortions}
\label{sec:results}

In \cref{sec:theory} we demonstrated qualitatively that disregarding the effect of AP distortions on void finding will likely lead to a biased result. Now we will quantify those biases using a template approach, as well as propose a couple of methods to mitigate the issue. 

%---------------------------------------

\subsection{Isotropic real-space CCF (template)}
\label{sec:results_template}

\begin{figure}
	\centering
	\includegraphics[width=0.85\columnwidth]{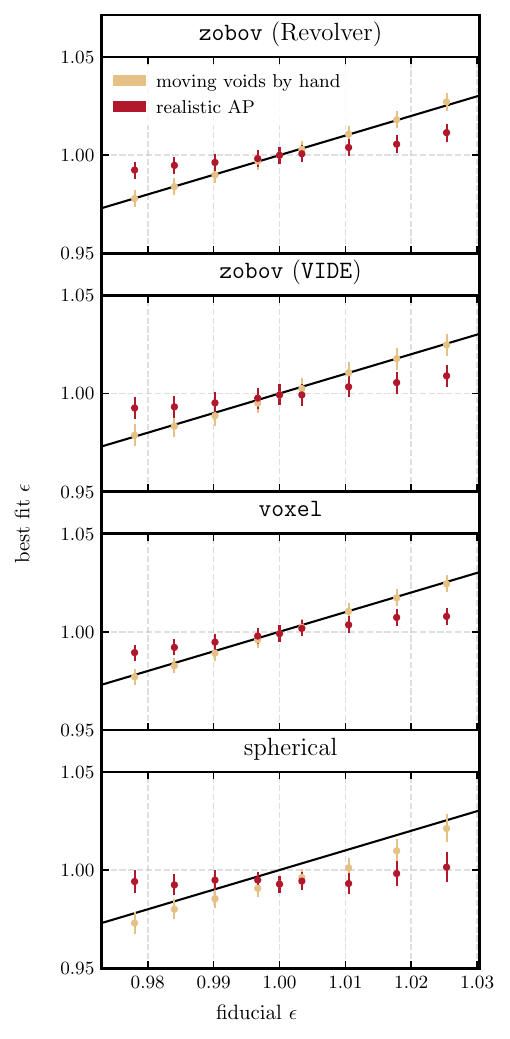}
	\caption{Constraints on $\epsilon$ when assuming different fiducial flat $\Lambda$CDM cosmologies (top part of \cref{tab:all_cases}). Yellow points are for the test case where both halos and voids were moved by hand, while the red points are for a more realistic case, where voids were found in the distorted halo field. The $x$-axis shows the $\epsilon$ corresponding to the fiducial cosmology, while the $y$-axis is the best fit. The diagonal black line represents the point where the best fit epsilon matches the fiducial one. }
	\label{fig:fit_assumed}
\end{figure}

We first constructed a scenario in which the assumption of void movement under AP is true by applying an AP shift to both halos and voids for 9 fiducial flat $\Lambda$CDM cosmologies (top part of \cref{tab:all_cases}). Using an isotropic template $\xir$ to calculate the theory, we obtained constraints on $\epsilon$ that are given in yellow in \cref{fig:fit_assumed}, where we see that $\epsilon$ is successfully recovered for all of the fiducial cosmologies and across almost all void finders\footnote{The spherical void finder shows a $1$ to $2\sigma$ offset even in this case, possibly due to the sharp features of spherical voids being sensitive to noise.}. This is show by the yellow points, which all following the black line for which the best fit $\epsilon$ is equal to the fiducial one, meaning that the true cosmology is successfully recovered. This means that the model indeed works well if the assumption on void movement under AP is true. 

Next we took a more realistic approach, allowing AP shifts in halo positions and performing void finding on this AP distorted halo catalogue, which is equivalent to the application of the analysis to real data. We again used a template $\xir$ to calculate the theory, like in the previous case, so the results are expected to match the previous ones if the assumption that voids move like tracers under AP is true. However, in \cref{fig:fit_assumed} (in red) we explicitly see that the results differ significantly and that this approach leaves us with a strong bias when the fiducial cosmology is different than the truth. While these results are unsurprising given \cref{fig:demonstration,fig:demonstration2}, here we see that not only are the results biased for all void finders, but this approach also strongly favours values of $\epsilon$ closer to one, regardless of the fiducial cosmology. 

This appears to be a direct result of the additional signal in the CCF seen in \cref{sec:theory}, which is induced by the void finder response to the AP distortion in tracer positions.  As the effect is most prominent in the quadrupole, we expect that any method of obtaining the real-space CCF which excludes this quadrupole will return a biased result. In this paper we present two ways that attempt to mitigate this issue without resorting to analytic modelling of the effect.

%---------------------------------------

\subsection{Approach 1: anisotropic real-space CCF}
\label{sec:results_fromdata}

\begin{figure*}
	
	\centering
	\subfloat{
	\includegraphics[width=0.425\hsize]{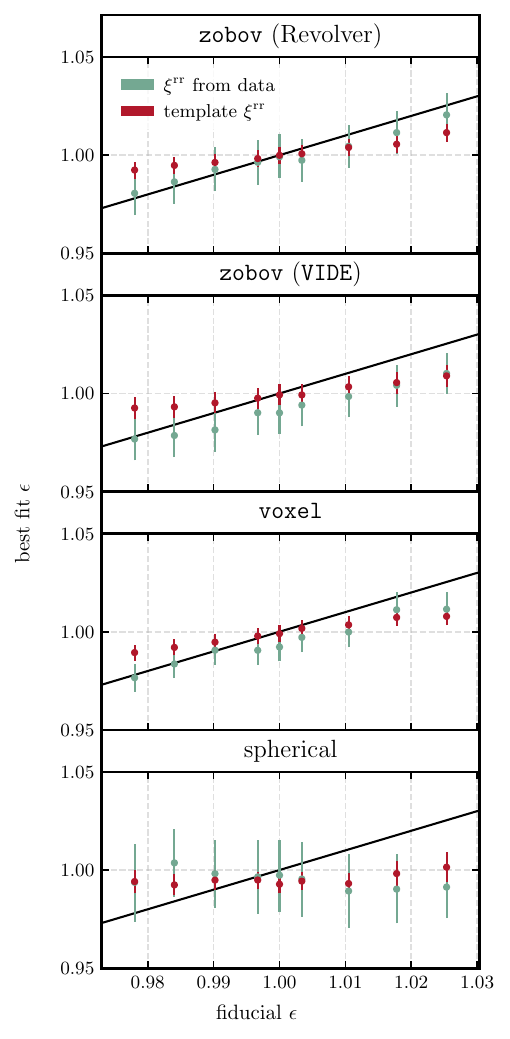} }
	\subfloat{
	\includegraphics[width=0.425\hsize]{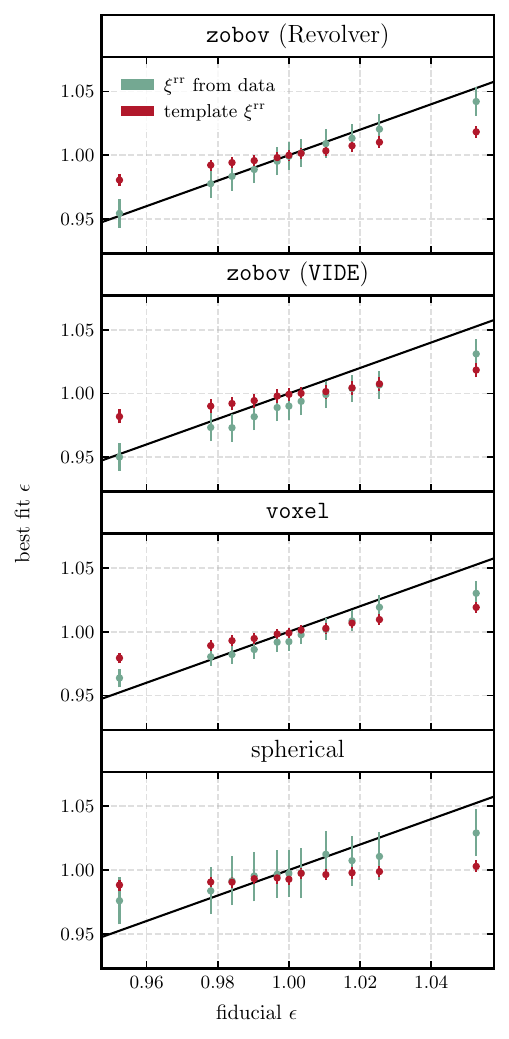} }
	
	\caption{Constraints on $\epsilon$ when assuming different cosmologies, flat $\Lambda$CDM on the left and fixed $q=1$ on the right. Red points are obtained when using a template $\xir$ (and match the red points in \cref{fig:fit_assumed}), while the green ones are obtained when $\xir$ is taken from data. The $x$-axis shows the $\epsilon$ corresponding to the fiducial cosmology, while the $y$-axis is the best fit. The diagonal black line represents the point where the best fit epsilon matches the fiducial one. }
	
	\label{fig:fit_cosmo}
\end{figure*}

Since the effect of AP distortions on void finding leaves an imprint in the real-space CCF that is not accounted for with an isotropic template CCF  (\cref{fig:demonstration}), we can circumvent the issue by taking the anisotropic real-space CCF $\xir$ directly from the data. For simplicity, in the current work, we measure the real-space CCF in the simulations, but note that this procedure is also generalisable to real data applications, where approximation to the real-space CCF can be obtained from the data using reconstruction, as demonstrated in \citet{Euclid:2023eom}. Whatever the case, this ensures that the additional effect will be included and can be propagated directly to the redshift-space CCF via the RSD mapping. Additionally, as both the model input $\xir$ and the data vector $\xis$ come from the same set of voids and halos, the model prediction and data will themselves be correlated and, to account for this, we use the full covariance matrix $\mathsf{C}\superscr{tot}$ as defined in \cref{eq:DMTcov}. 

Because the real-space CCF is measured in the fiducial cosmology, we do not perform any AP rescaling on it. Instead, the only part of \cref{eq:streaming_model} that needs to be rescaled is the velocity PDF, which is in the reference cosmology. This means that the constraining power on $\epsilon$ is greatly reduced, since the AP distortion on $\xir$ is not being modelled, and we are only getting constraints from the effect of AP distortion to the velocity PDF. The result of applying this approach to all the fiducial cosmologies given in \cref{tab:all_cases} is shown in green, in \cref{fig:fit_cosmo}, alongside the results from the previous section (red). Though the constraint on $\epsilon$ is weaker, we are closer to recovering the fiducial values, in some cases even for cosmologies very far from the true one (extreme points in the right panel of \cref{fig:fit_cosmo}). The \revolver implementation of \zobov in particular responds well to this method, recovering the correct value of $\epsilon$ to within $1\sigma$ in all cases. In general, all of the void finders seem to handle isovolumetric distortion well, perhaps due to the velocity templates being better suited for those cases. This is particularly obvious for the spherical void finder, which does better only for isovolumetric distortions, while the flat $\Lambda$CDM cosmologies remain seemingly insensitive to $\epsilon$. \voxel and \vide show mixed results, performing poorly for fiducial $\epsilon$ values much greater than one, but well otherwise, with \voxel performing a little better than \vide. This indicates that higher values of $\Omega\subscr{m}\superscr{fid}$ are a safer assumption for these void finders. 

We can think of AP distortions to the CCF depending on shifts in both void and tracer positions, which are described in \cref{sec:APD}, and a more complicated effect, which is a function of the specific void-finding algorithm used, and results in the differences we see in \cref{fig:demonstration,fig:demonstration2}. The success of the approach presented here suggests that these effects are at least somewhat separable; rescaling the CCF using AP scaling parameters accounts for the AP shifts, while the effect on void finding is left to propagate through the RSD mapping, and is correctly included in the calculated redshift-space CCF. 

%---------------------------------------

\subsection{Approach 2: Iterative template-based modelling}
\label{sec:results_FM}

\begin{figure*}
	\centering
	\includegraphics[width=1.8\columnwidth]{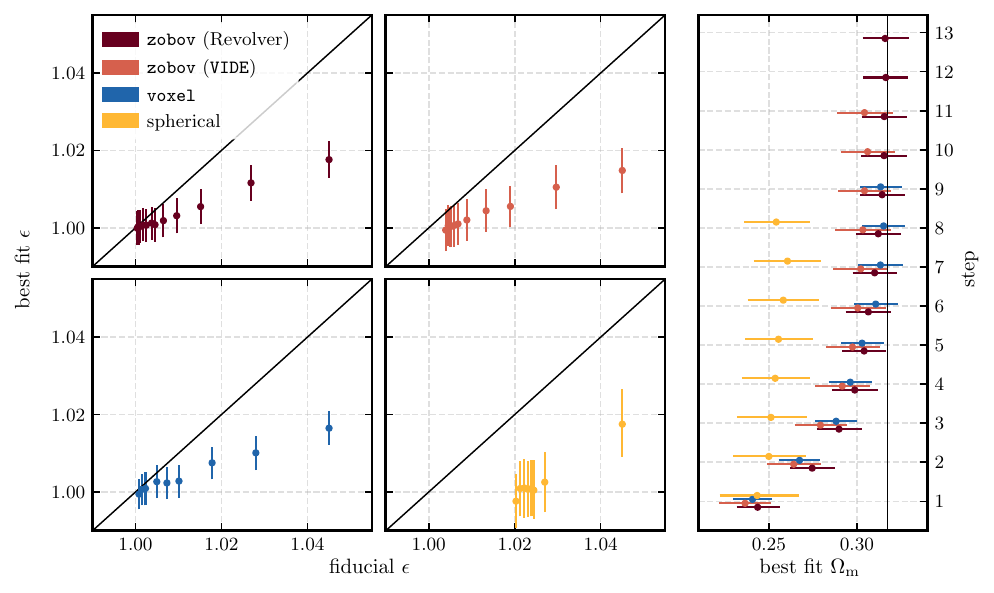}
	\caption{Constraints on $\epsilon$ and $\Omega\subscr{m}$ obtained using iterative template-based modelling. The four panels on the left show best fit $\epsilon$ values obtained at each step, while the right panel shows the resulting $\Omega\subscr{m}$ constraints, assuming a flat $\Lambda$CDM universe. The black vertical line is the true $\Omega\subscr{m}$ of the Quijote simulations, while the four diagonal black lines represent the points where best fit epsilon matches the fiducial one.}
	\label{fig:FMA_template}
\end{figure*}

While the first approach seems to work well, it does result in weaker constraints, and depends strongly on our ability to accurately obtain the real-space CCF using, for example, reconstruction. An alternative presents itself in \cref{fig:fit_assumed}, where we see that, although the red points are biased, the bias itself depends on $\epsilon$. Iterating using the template approach, we can try to recover the true cosmology step by step.

We start with a flat $\Lambda$CDM fiducial cosmology with $\Omega\subscr{m}\superscr{fid}=0.2$, which corresponds to $\apar=0.9170$ and $\aperp=0.9583$. We stretch the halo positions according to those values, and run all of our void-finding algorithms to obtain the void catalogues and, finally, data vectors. The real-space CCF and velocity profiles are supplied as templates, like in \cref{sec:results_template}. This gives us a constraint on $\epsilon$ that can be translated into $\Omega\subscr{m}$, which will be our new fiducial cosmology. From our new fiducial $\Omega\subscr{m}\superscr{fid}$, we can in turn get new rescaling parameters using \cref{eq:AP_apar_dist,eq:AP_aperp_dist}, and the true value $\Omega\subscr{m}\superscr{true}$ from the simulation. Because of the bias we find in \cref{sec:results_template}, it is not enough to recover a value of $\epsilon$ that is consistent with $1$ to within $1$--$2\sigma$, we have to make sure we are actually close to the truth. Therefore, this process is repeated until the best-fit value of $\epsilon$ crosses the point of $\epsilon=1$. 

The results at each step are given in \cref{fig:FMA_template}, with constraints on $\epsilon$ in the four left-most panels and constraints on $\Omega\subscr{m}$ in the right column. For $\epsilon$ we aim to recover back the fiducial value of $\epsilon$ that we put in by assuming a cosmology and stretching the simulation boxes accordingly. In essence, the $x$-axis represents our fiducial cosmology, and the $y$-axis tells us if that guess is correct. The spherical void finder, for example, very quickly returns $\epsilon=1$ even when our fiducial cosmology is very far from the truth. This would, in turn, result in a wrong estimate of $\Omega\subscr{m}$, as we see in \cref{fig:FMA_template} (right panel). While choices like radius binning might affect the result to some degree, the spherical void finder seems inadequate for this specific approach, possibly due to the sharp features which are sensitive to noise. 

The \revolver implementation of \zobov, on the other hand, gives an unbiased result but takes a lot of steps to reach it, while \vide takes slightly fewer steps but results in more bias. Moreover, the biggest issue for both is that void finding itself is very time-consuming for tessellation based methods. A good compromise between speed and accuracy for this approach is the \voxel void finder, which manages to quickly recover the correct value of $\Omega\subscr{m}$ to within $1\sigma$, while running about 10 times faster than \zobov. However, the difference in accuracy between the two \zobov void finders might provide an interesting insight into the issue, since the void-finding step is the same, with the only difference being in the centre definition: \revolver defines the centre using only four particles, while \vide employs a more complicated approach, considering all of the particles belonging to the void. The point-based centre of \revolver voids might be better equipped to follow the AP shifts of galaxies, unlike voids with non-local centre definitions. Furthermore, it is likely that it is the density estimation step (which all of the void finders here employ) that 'smears out' the galaxy, and thus void, positions and affects the expected AP shifts. A point-based void-finding algorithm, which works directly with galaxy positions without estimating the density field, could be an interesting avenue to follow in this case. 

In general, the success of each void finder in this approach can be deduced already from the red points in \cref{fig:fit_cosmo}, since they trace the possible path that the iterative approach can take. Thus, testing the suitability of new, different void finders to this approach can be done by repeating the analysis in \cref{sec:results_template}.

%#######################################################

\section{Conclusions}
\label{sec:conclusions}

In this paper, we have investigated the effect of AP distortions on the void-galaxy cross-correlation function. Previous studies relied on an implicit assumption that the process of void finding is unaffected by AP distortions, with void positions in a given fiducial cosmology simply related by an AP shift to void positions in the true cosmology, similarly to galaxies. We have shown that this assumption is generally incorrect, especially when the fiducial cosmology differs significantly from the true one. This finding has important implications for the modelling of the void-galaxy CCF and the results of previous studies of the AP effect in voids that have recovered constraints on the AP distortion parameter $\epsilon$. Although we mainly focused on template approaches to the CCF modelling in this study, we stress that this effect impacts all theoretical models of the void-galaxy CCF, and is present regardless of whether voids are located in real or redshift space. Despite this, the constraining power of voids, i.e. the error on $\epsilon$, is not likely to be affected if the iterative approach is used. Depending on void finder, previous constraints that recover $\epsilon = 1$, i.e. where the chosen fiducial cosmology is very close to the true one, will be less affected; however, as seen in \cref{fig:fit_cosmo}, this is not a guarantee that biases are not present. In particular, some previous studies \citep{Nadathur:2020,Aubert20a,Woodfinden:2022} have investigated the systematic error related to the choice of fiducial cosmology, and correctly accounted for this bias by including it into the total error budget. 

A more appropriate treatment of this effect would require us to develop a host of models for all of the different void finders, since the effect itself depends on the specific algorithm used. Alternatively, one could adopt a machine learning approach, in which a neural network could be trained to recognize the finder response to AP \citep[in prep.]{Fraser:2024}. Though this approach has proven to work well, it is also finder specific, as well as computationally intensive in its initial stages. In this paper, we instead proposed two methods that enable us to mitigate the effect of void finder response to AP, while avoiding complicated modelling. 

Since the effect of AP distortions on void finding leaves an imprint in the real-space CCF, that is not accounted for with an isotropic CCF, we can circumvent the issue by estimating the real-space CCF directly form the data. This real-space CCF contains the anisotropies induced by the fiducial cosmology, helping us propagate the unknown contribution from the void finder response and recover unbiased cosmological parameters. This way we can recover the true $\epsilon$ values even when the fiducial cosmologies are very far from the true one. The downsides of this approach, however, are that it results in weaker constraints, and depends heavily on our ability to accurately obtain the real-space CCF (using e.g. reconstruction).

The second method we have investigated is an iterative approach. We perform a full analysis using a fixed fiducial cosmology and, using the recovered value of $\epsilon$, we redefine the fiducial cosmology (i.e. $\Omega\subscr{m}$) iteratively until the new best fit value crosses the point of $\epsilon=1$ (meaning an $\epsilon>1$ becomes $\epsilon<1$, or vice versa). We found that this approach works well in most cases, but depends a lot on the void finder in question. The spherical void finder converges fast, but gives a very biased result; \zobov gives an unbiased result, but is very slow to converge; while the \voxel void-finder is both fast to converge and gives unbiased results. A particular benefit of the iterative approach is that it can be easily applied to previous studies, to ensure that the recovered parameters are unbiased, without sacrificing the constraining power.

%#######################################################

\section{Acknowledgements}
%Sladja 
SR acknowledges support from the Research Council of Norway through project 325113. 
% Sesh
SN acknowledges support from an STFC Ernest Rutherford Fellowship, grant reference ST/T005009/2. 
% Elena, Enrique and Will
Research at Perimeter Institute is supported in part by the Government of Canada through the Department of Innovation, Science and Economic Development Canada and by the Province of Ontario through the Ministry of Colleges and Universities. 
% open access
For the purpose of open access, the authors have applied a CC BY public copyright licence to any Author Accepted Manuscript version arising.

%#######################################################

\bibliographystyle{aa}
\bibliography{references}

%#######################################################

\appendix

\section{Data vectors}
\label{app:datavectors}

\begin{figure*}
	\centering
	\includegraphics[width=1.8\columnwidth]{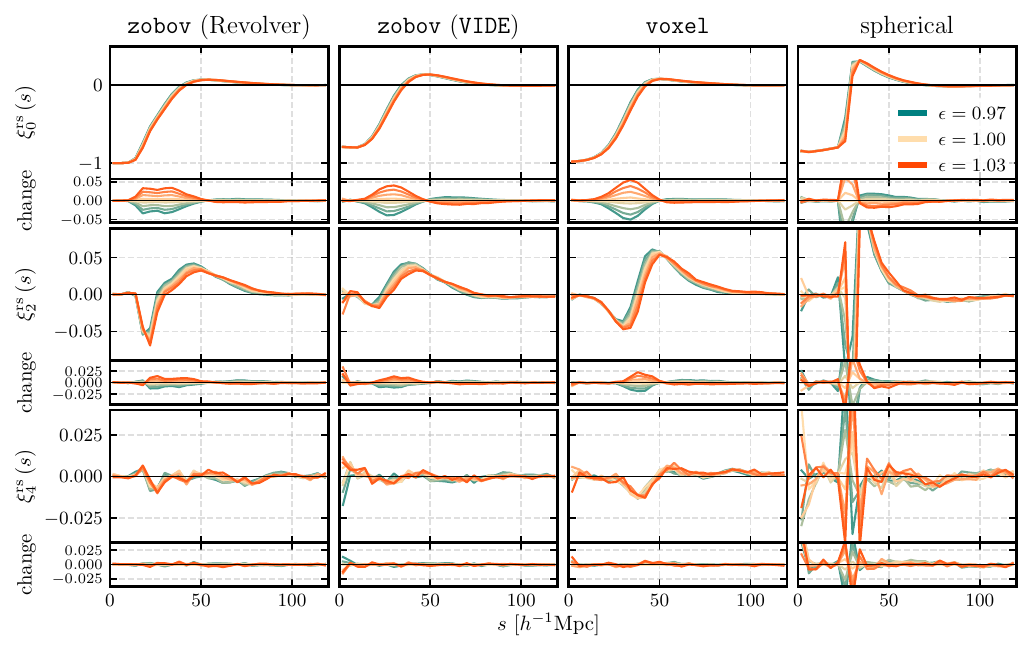}
	\caption{Data vectors for all of the void finders for different flat $\Lambda$CDM fiducial cosmologies (top part of \cref{tab:all_cases}). We see that the hexadecapole is almost completely noise dominated, while the monopole and quadrupole show a clear response to the change in $\epsilon$ and $q$. }
	\label{fig:datavectors_cosmo}
\end{figure*}

\begin{figure*}
	\centering
	\includegraphics[width=1.8\columnwidth]{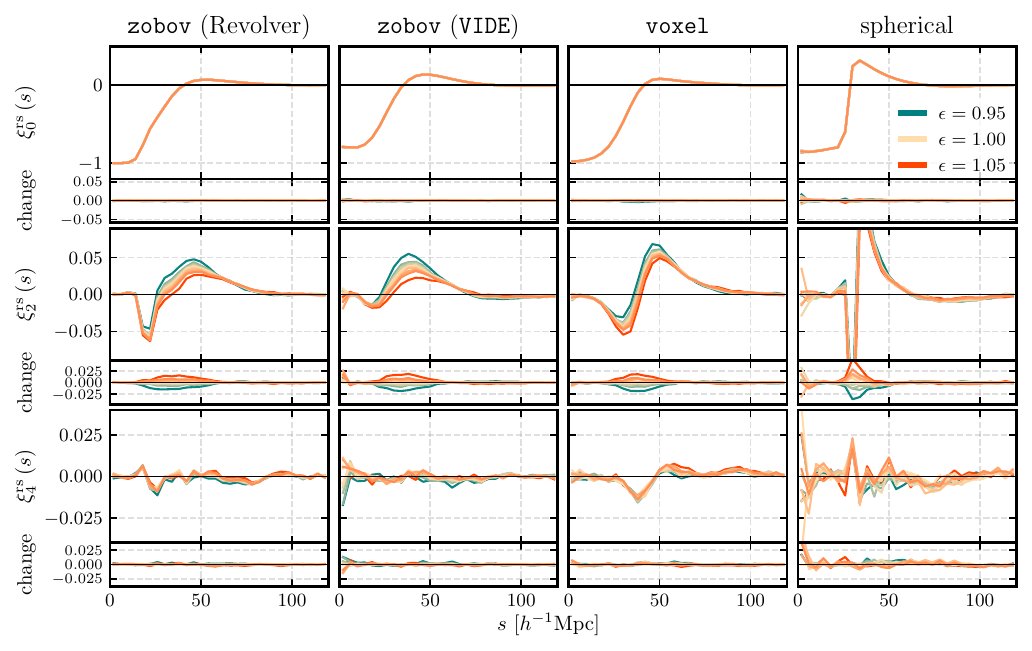}
	\caption{Data vectors for all of the void finders for different fiducial cosmologies with isovolumetric distortions ($q=1$, lower part of \cref{tab:all_cases}). While the quadrupole shows a strong response to the change in $\epsilon$, the monopole does not change significantly since the void size stays constant in the different cosmologies. }
	\label{fig:datavectors_epsilon}
\end{figure*}

For completeness, we show here all of the data vectors measured for the analysis presented in this paper. They cover all of the fiducial cosmologies listed in \cref{tab:all_cases}: the nine flat $\Lambda$CDM cosmologies are shown in \cref{fig:datavectors_cosmo} and the 12 cosmologies with isovolumetric AP distortion are shown in \cref{fig:datavectors_epsilon}. The data vectors for all of the four void finders are comprised of a monopole, quadrupole, and hexadecapole moment, with the hexadecapole being largely dominated by noise. An exception to this is the \voxel void finder, which seems to show a clear signal in the hexadecapole, even though it does not vary with $\epsilon$. 

\Cref{fig:datavectors_epsilon} shows that when $q$ is kept fixed, there is no significant change in the monopole, while the quadrupole contains the bulk of the information on the changing distortion parameter $\epsilon$. This is expected as the monopole is mostly sensitive to the change in the overall volume of the void, while a changing $\epsilon$ only distorts the shape, which is described by the quadrupole. Indeed, in \cref{fig:datavectors_cosmo}, where $\epsilon$ and $q$ both change, we see that the change in the overall void volume (or size) is evident in both the monopole and multipole. 

%---------------------------------------

\section{Voids in redshift space}
\label{app:redshiftspace}

\begin{figure*}
	\centering
	\subfloat{ \includegraphics[width=0.425\hsize]{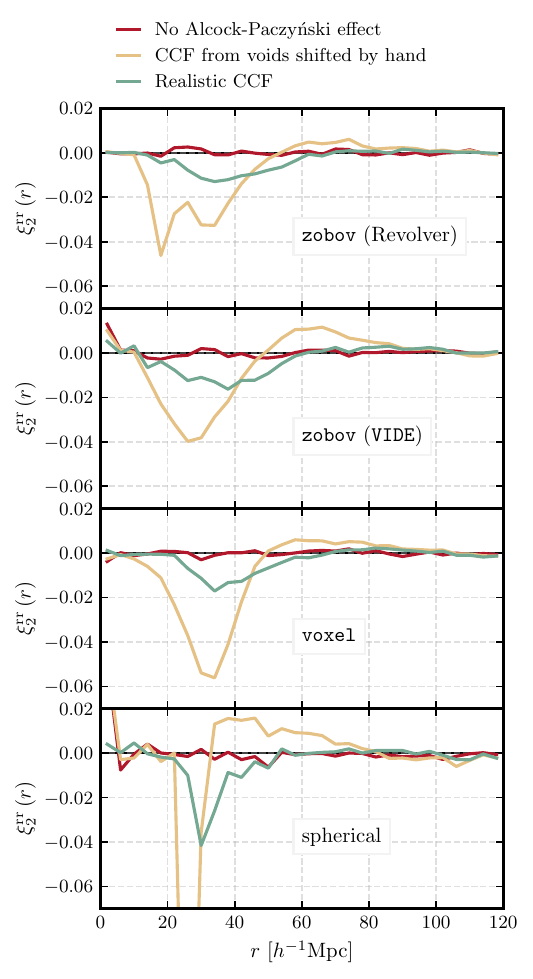} }
	\subfloat{ \includegraphics[width=0.425\hsize]{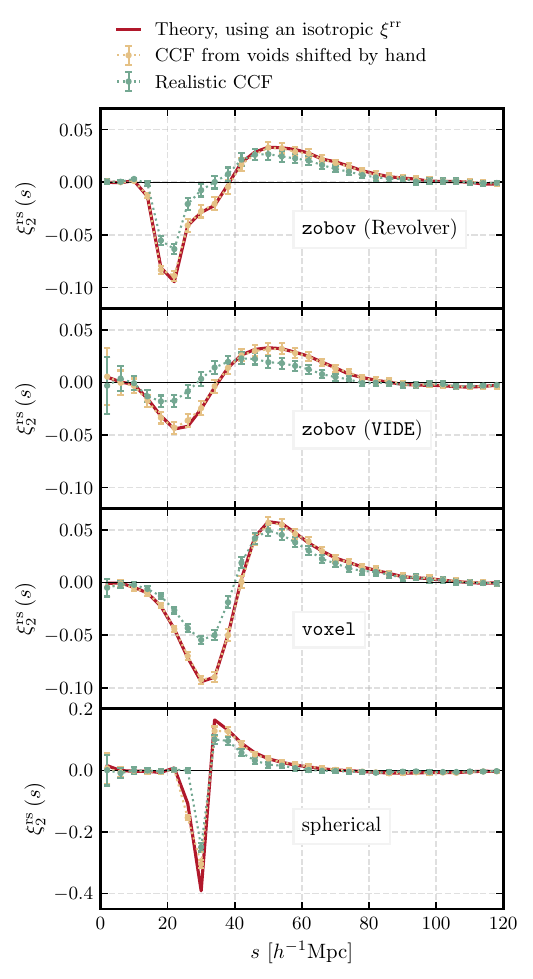} }
	\caption{Quadrupole moments of the real and redshift-space CCFs for all void finders, with voids found in real space. 
	\textit{Left:} Quadrupoles of the real-space CCF in the true cosmology (red) and a fiducial cosmology with $\epsilon=1.053$ and $q=1$ (yellow and green). The yellow line is the assumed case where voids move under AP distortion in the same way as halos, while the green line is the more realistic case where voids are found in the distorted halo field. 
	\textit{Right:} Quadrupoles of the redshift-space CCF in a fiducial cosmology which differs from the truth ($\epsilon=1.053$ and $q=1$). In red we plot the theory prediction, calculated using a template $\xir$.}
	\label{fig:abs_all_finders}
\end{figure*}

\begin{figure*}
	\centering
	\subfloat{ \includegraphics[width=0.425\hsize]{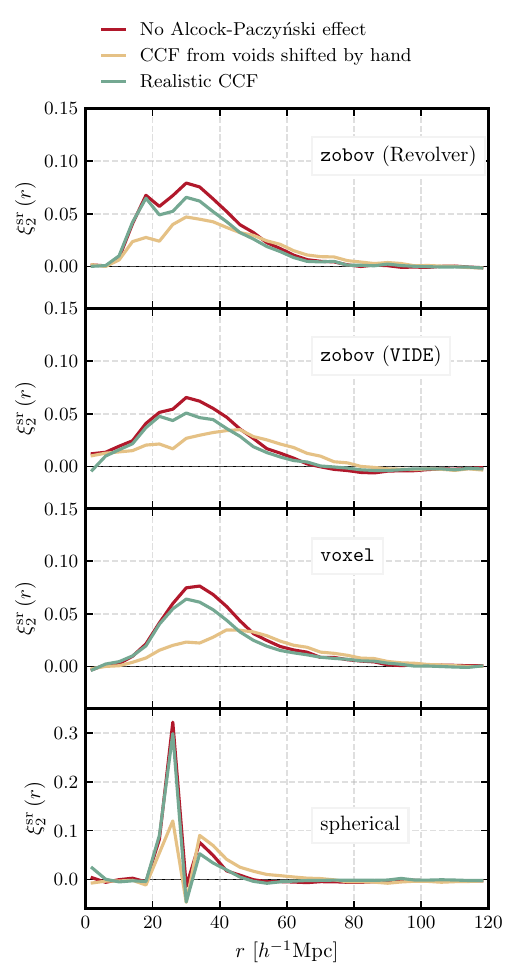} }
	\subfloat{ \includegraphics[width=0.425\hsize]{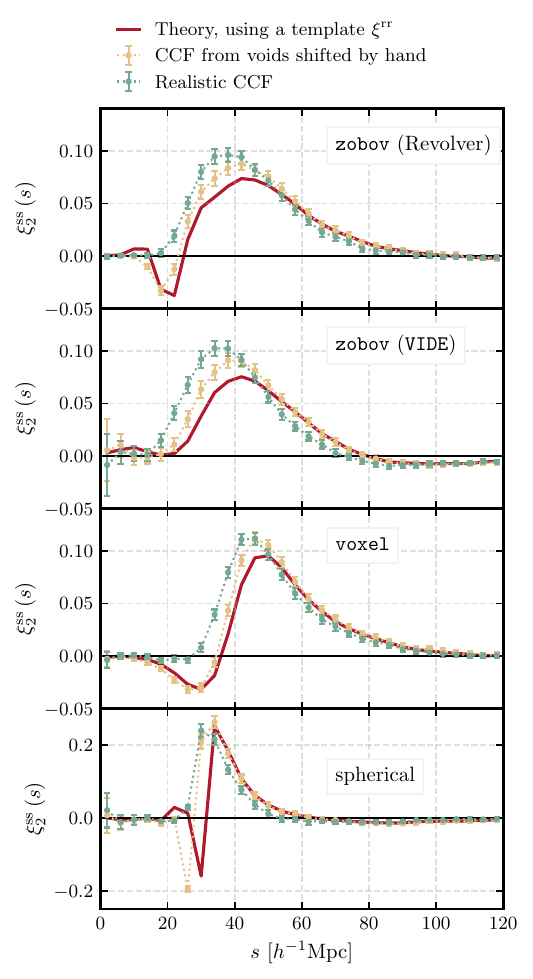} }
	\caption{Quadrupole moments of the real and redshift-space CCFs for all void finders, with voids found in redshift space. 
	\textit{Left:} Quadrupoles of the real-space CCF in the true cosmology (red) and a fiducial cosmology with $\epsilon=1.053$ and $q=1$ (yellow and green). The yellow line is the assumed case where voids move under AP distortion in the same way as halos, while the green line is the more realistic case where voids are found in the distorted halo field. 
	\textit{Right:} Quadrupoles of the redshift-space CCF in a fiducial cosmology which differs from the truth ($\epsilon=1.053$ and $q=1$). In red we plot the theory prediction, calculated using a template $\xir$.}
	\label{fig:abs_all_finders_redshiftspace}
\end{figure*}

In this appendix we show some additional results for the idealised test performed in \cref{sec:APvoids}. While in \cref{sec:APvoids} we demonstrate the effect only for one of the void finders, here we show that it is an effect that persists in all of the algorithms that we investigate, excluding the possibility of a finder-specific error. In \cref{fig:abs_all_finders} we show the quadrupoles of the real-space CCF we get when we assume voids move under AP distortion in the same way as halos compared to the case where voids are found in the distorted halo field, corresponding to the realistic scenario. The results given in \cref{fig:demonstration,fig:demonstration2} are in the third panel (\voxel), but we see that the other void finders show the same trend, namely that AP distortions have smaller amplitude in reality (green) than expected (yellow). 

The results in \cref{sec:APvoids} all have voids identified in real space, but void-finding can be run on the redshift-space catalogue as well. In this case the tracer positions contain RSD as well, introducing another source of anisotropy into the void shapes. This is evident in the left panel of \cref{fig:abs_all_finders_redshiftspace}, where the quadrupole of the real-space CCF (now $\xi^{\rm sr}$ instead of $\xir$) in the true cosmology is no longer zero and instead shows that voids are stretched along the line of sight. Despite these differences, we still see a large difference between the expected quadrupole in the fiducial cosmology (yellow, assuming voids move under AP in the same way as tracers) and the realistic one (green), showing that the effect also persists regardless of what space the voids are located in. In the right panel we again plot the measured redshift-space CCFs, as well as the model calculations. It is worth noting that, even in the idealized case, the model does not describe the data well, because there is a selection effect due to the voids having a preferred orientation, similarly to the voids in a fiducial cosmology with AP distortions. Unlike the AP, however, this selection effect can not be attenuated or circumvented by the methods described in this paper.

%#######################################################

\end{document}